# Thermodynamics of the inner heliosheath

G. Livadiotis [1], D.J. McComas [1], H. O. Funsten [2], N.A. Schwadron [1,3], J. R. Szalay[1], E. Zirnstein [1]

[1] Department of Astrophysical Sciences, Princeton University, Princeton, NJ 08544, USA; glivadiotis@princeton.edu
[2] Los Alamos National Laboratory, ISR Division, P.O. Box 1663, Los Alamos, NM 87545, USA
[3] University of New Hampshire, Space Science Center, Morse Hall, Durham, NH 03824, USA

**Abstract**

We derive annual sky-maps of the proton temperature in the inner heliosheath (IHS), and track their temporal evolution over the years 2009–2016 of Interstellar Boundary Explorer observations. Other associated thermodynamic parameters also determined are the density, kappa - the parameter that characterizes kappa distributions, temperature rate, polytropic index, and entropy. We exploit the theory of kappa distributions and their connection with polytropes, to (i) express a new polytropic quantity $\Pi$ that remains invariant along streamlines where temperature and density may vary, (ii) parameterize the proton flux in terms of the $\Pi$ invariant and kappa, and (iii) derive the temperature and density, respectively, from the slope and intercept of the linear relationship between kappa and logarithm of $\Pi$. We find the following thermodynamic characteristics: (1) Temperature sky-maps and histograms shifted to their lowest values in 2012 and their highest in 2015; (2) Temperature negatively correlated with density, reflecting the sub-isothermal polytropic behavior; (3) Temperature positively correlated with kappa, revealing characteristics of the mechanism responsible for generating kappa distributions; (4) Processes in IHS are sub-isothermal tending toward isobaric, consistent with previously published results; (5) Linear relationship between kappa and polytropic indices, revealing characteristics of the particle potential energy; and (6) Entropy positively correlated with polytropic index, aligned with the underlying theory that entropy increases towards the isothermal state where the kappa distribution reduces to the Maxwell-Boltzmann description.

## 1. Introduction

The Interstellar Boundary Explorer (IBEX) mission was launched in late 2008 and is dedicated to measure Energetic Neutral Atoms (ENAs) from the outer heliosphere. Analysis of these measurements enables understanding of the interaction between the solar wind and the very local interstellar medium (VLISM) (McComas et al. 2009a). A special issue of Science in 2009 published the IBEX first results (Funsten et al. 2009a; Fuselier et al. 2009a; McComas et al. 2009b; Möbius et al. 2009; Schwadron et al. 2009). By the end of 2021, IBEX had captured a complete set of observations over the full solar cycle 24 (McComas et al. 2020).

ENAs are produced by charge exchange between energetic ions and cold interstellar neutral atoms drifting in from the VLISM. Heliospheric ENA observations are divided into to two broad sources: (1) the globally distributed flux (GDF) (McComas et al. 2009b; Schwadron et al. 2011) encompassing the entire sky with broad enhancements and depletions in different directions; and (2) the narrow "ribbon" of enhanced ENA emission that encircles the sky (McComas et al. 2009b; 2010; Schwadron et al. 2009; Funsten et al. 2009a; 2013; 2015; Fuselier et al. 2009a; Swaczyna et al. 2016; Schwadron & McComas 2019). This paper focuses on GDF emissions, whose source is predominantly the inner heliosheath (IHS), where the suprathermal pickup proton component of the solar wind is preferentially heated at the termination shock (e.g., Zank et al. 1996, 2010; Richardson et al. 2008; Richardson & Stone 2009; Yang et al. 2015; Kumar et al. 2018) and dominates the one to few keV ENA GDF signal.

IBEX acquires all-sky ENA flux maps using two single pixel imagers that accumulate a circular swath of data each spin. This circular swath moves over the sky as the IBEX spin axis repoints to the Sun throughout the year (see: IBEX-Hi, Funsten et al. 2009b, IBEX-Lo, Fuselier et al. 2009b). This results in 1800-pixeled sky maps of the ENA flux every ∼6 months, covering a broad energy range from ∼0.01 to ∼6 keV.

For this study, the proton fluxes used to produce the heliosheath proton distribution properties were calculated by Zirnstein et al. (2021a) using annual IBEX-Hi observations taken in the spacecraft's ram reference frame. The ram frame (McComas et al. 2012) corresponds to measurements taken when the spacecraft was moving with some velocity in the direction of the Line-Of-Sight (LOS) viewing.

Using IBEX-Hi ENA observations from 2009 with energies between ∼0.5 and 6 keV, Livadiotis et al. (2011;2012;2013) derived the sky maps of the values of temperature and other thermodynamic quantities of the ENA-source proton plasma in the IHS. This was accomplished by connecting the observed ENA flux to the energy kappa distribution of the source protons.

More specifically, the kappa distribution is used to parameterize the proton population in the IHS, because, it (1) is consistent with the power-law behavior observed in ENA spectra (Livadiotis & McComas 2009, 2010; Livadiotis et al. 2011); (2) describes space plasmas throughout the heliosphere (e.g., Livadiotis & McComas 2009; 2013a; Tsallis 2009; Livadiotis 2015a; 2017; and refs. therein); (3) has strong foundations within the context of nonextensive statistical mechanics (Treumann 1997; Milovanov & Zelenyi 2000; Leubner 2002; Livadiotis & McComas 2009; 2010; Livadiotis 2014a; 2015a;b; 2017, Ch.1; 2018a); (4) describes the polytropic behavior of space plasma flows (e.g., Nikolaou & Livadiotis 2019; Livadiotis 2019a); and, (5) leads to a consistent characterization of temperature for systems residing in stationary states out of thermal equilibrium, such as the majority of space plasmas (Livadiotis & McComas 2009; 2010; 2021; Livadiotis 2018b).

The proton energy-flux ($\varepsilon$, $j$) spectra, deduced from ENA measurements, are characterized by a power-law behavior with uncertainties well described by a log-normal distribution. Hence, a linear relationship is used for plotting and analyzing the spectra on a log-log scale, (log $j$ vs. log $\varepsilon$); the linear fit determines the intercept of the flux logarithm, log $j_*$ (where $j_*$ is the flux at 1 keV), and the spectral index, $\Gamma$. There is another important linear relationship; this is between the fitting parameters of spectra, i.e., the flux-logarithm intercept log $j_*$ and the kappa $\kappa$; the latter constitutes the parameter that labels and characterizes the kappa distributions and can be well approached by the spectral index $\Gamma$. This linearity between the flux-logarithm intercept and kappa has been noted over the past decades (e.g., Schreier et al. 1971; Crosby et al. 1993; Von Montigny et al. 1995); it is important, because the two fitting parameters ($\kappa$, log $j_*$) can provide the temperature and density, and the characterization of the thermodynamics.

In the earlier analyses of ENA observations from 2009, we exploited the above linearity between $\kappa$ and log $j_*$ to derive the temperature $T$ and density $n$ of each sky-map pixel. In particular, we applied a linear regression to the ENA flux spectrum using the expression, i.e., log $j(\varepsilon)$ = log $j_*$ – $\kappa$ log $\varepsilon$, from which we derived the kappa $\kappa$ and the intercept log $j_*$. The fitting of the energy-flux spectrum generates one pair of ($\kappa$, $j_*$) values; however, more than one pairs is needed for the second linear fitting to derive the temperature and density. A special statistical technique (described in Livadiotis et al. 2011, 2012; 2013; Livadiotis 2016) was used for generating a statistically significant number of such pairs. The linear fitting of the derived pairs of ($\kappa$, log $j_*$) values, leads to the estimation of the values of the temperature $T$ and density $n$, derived respectively from the slope and intercept of this linear regression.

While this method was self-consistently applied to each pixel, significant uncertainties come along with the temperatures and their variation over the various sky-map directions. A new method is applied in this study, where the temperature and density are determined by smoothing with the values of ($\kappa$, $j_*$) over neighboring pixels. In particular, we show, first, that a polytropic quantity $\Pi$ exists, which remains invariant despite the variations of the temperature and density along the flow streamline. The $\Pi$ invariant is proportional to the flux argument $j_*$; thus, the proton flux is parameterized in terms of ($\kappa$, $\Pi$). We show, then, that the observed values of ($\kappa$, log $\Pi$), similar to ($\kappa$, log $j_*$), exhibit a linear relationship, where $T$ and $n$ are again derived respectively from the slope and intercept of this linear regression.

The method for deriving the temperature and density is based on two types of linear fits: First, the linear fit of the energy-flux spectrum of each sky-map pixel leading to a pair of ($\kappa$, log$\Pi$) values. Then, combining a number of neighboring pixels, we produce a number of ($\kappa$, log$\Pi$) pairs, whose linear fit provides the values of temperature and density, smoothed over this selection of neighboring pixels. There are eight neighboring pixels immediately surrounding each pixel, so these nine pixels can provide a sufficient number of pairs of ($\kappa$, log$\Pi$) for the second linear fitting. Moreover, an alternative technique is preferred for performing the second fitting: we first combine the examined pixel with each of the eight neighboring pixels, and determine the values of temperature and density. We derive eight values of temperature and density, one for each of the eight combinations of neighboring pixels with the examined central pixel; and then, we find their weighted averages and uncertainties. The latter are the outcome of two independent types of errors, the fitting errors and the propagation errors, which are based on the fits involved in the proton flux spectrum and the propagation of the flux spectrum errors.

The purpose of this study is to analyze and interpret the proton energy-flux spectra of the GDF emissions of heliospheric ENAs over the years 2009-2016, in order to derive the temperature and other thermodynamic parameters, and finally, characterize the thermodynamic state of the proton plasma in the IHS; these parameters include the density, kappa, polytropic index, and entropy.

The paper is organized as follows. Section 2 presents the theory of kappa distributions, the connection to polytropic behavior of plasmas, and the applications to this study. We show: (i) the formulation of proton flux with respect to its energy, (ii) that the thermodynamic parameter kappa $\kappa$ that characterizes the kappa distributions and the polytropic invariant quantity $\Pi$ that characterizes the polytropic processes, are the only two parameters involved in the energy-flux spectra; and (iii) the dependence of $\Pi$ on the temperature and density, and how they can be deduced from the values of ($\kappa$, $\Pi$). Section 3 describes the method for deriving temperature and density, through three statistical analyses, (i) fitting of spectra; (ii) smoothing of temperature and density values from any two neighboring pixels; and (iii) averaging of the smoothed values derived from all neighboring pixels. We also perform analyses for verifying the method, i.e., (i) the kappa formulation involved in energy-flux spectra; (ii) the high-energy limit, reducing the kappa distributions to power-laws; and (iii) the constancy of streamline density and temperature for neighboring pixels. Section 4 presents the results, i.e., (i) sky-maps of temperature, density, and kappa, (ii) annual histograms, and (iii) evolution of the histogram. Section 5 presents the derivation of associated thermodynamic parameters, i.e., the polytropic index, entropy, and temperature rate. Section 6 provides the conclusions. Finally, the appendices provide details of the concepts used in this study.

## 2. Flux of a kappa distributed population – (κ,Π)-dependence

In the presence of particle interactions, the kappa distribution function $f(\mathbf{r},\mathbf{u})$ (Appendix A) is:

$$f[\mathbf{u};n(\mathbf{r}),T(\mathbf{r}),\kappa] = \pi^{-\frac{3}{2}} \cdot (\kappa-\tfrac{3}{2})^{-\frac{3}{2}} \frac{\Gamma(\kappa+1)}{\Gamma(\kappa-\tfrac{1}{2})} \cdot n(\mathbf{r}) \cdot [\tfrac{2}{m}k_{\mathrm{B}}T(\mathbf{r})]^{-\frac{3}{2}} \cdot \left[1 + \frac{1}{\kappa-\tfrac{3}{2}} \cdot \frac{\tfrac{1}{2}m|\mathbf{u}-\mathbf{u}_{\mathrm{b}}|^{2}}{k_{\mathrm{B}}T(\mathbf{r})}\right]^{-\kappa-1}, \quad (1)$$

where $|\mathbf{u}-\mathbf{u}_{\mathrm{b}}| = \sqrt{u^{2}+u_{\mathrm{b}}^{2}-2u\cdot u_{\mathrm{b}}\cos(\omega)}$; $\omega$ defines the angle between the particle $\mathbf{u}$ and bulk $\mathbf{u}_{\mathrm{b}}$ velocity vectors. The kappa distribution, expressed in terms of the particle kinetic energy in an inertial frame of the heliosphere, $\varepsilon = \tfrac{1}{2}mu^{2}$ (also setting $\varepsilon_{\mathrm{b}} = \tfrac{1}{2}mu_{\mathrm{b}}^{2}$), becomes

$$f[\varepsilon;n(\mathbf{r}),T(\mathbf{r}),\kappa] = \pi^{-\frac{3}{2}} \cdot (\kappa-\tfrac{3}{2})^{-\frac{3}{2}} \frac{\Gamma(\kappa+1)}{\Gamma(\kappa-\tfrac{1}{2})} \cdot n(\mathbf{r}) \cdot [\tfrac{2}{m}k_{\mathrm{B}}T(\mathbf{r})]^{-\frac{3}{2}} \cdot \left[1 + \frac{\varepsilon + \varepsilon_{\mathrm{b}} - 2\sqrt{\varepsilon}\cdot\sqrt{\varepsilon_{\mathrm{b}}}\cos[\omega(\mathbf{r})]}{(\kappa-\tfrac{3}{2})k_{\mathrm{B}}T(\mathbf{r})}\right]^{-\kappa-1}. \quad (2)$$

The flux of a kappa distributed population is given by $j[\varepsilon;n(\mathbf{r}),T(\mathbf{r}),\kappa] = 2m^{-2} \cdot f[\varepsilon;n(\mathbf{r}),T(\mathbf{r}),\kappa] \cdot \varepsilon$, or

$$j[\varepsilon;n(\mathbf{r}),T(\mathbf{r}),\kappa] = 2m^{-2}\pi^{-\frac{3}{2}} \cdot (\kappa-\tfrac{3}{2})^{-\frac{3}{2}} \frac{\Gamma(\kappa+1)}{\Gamma(\kappa-\tfrac{1}{2})} \cdot n(\mathbf{r}) \cdot [\tfrac{2}{m}k_{\mathrm{B}}T(\mathbf{r})]^{-\frac{3}{2}} \left[1 + \frac{\varepsilon+\varepsilon_{\mathrm{b}}-2\sqrt{\varepsilon}\sqrt{\varepsilon_{\mathrm{b}}}\cos[\omega(\mathbf{r})]}{(\kappa-\tfrac{3}{2})k_{\mathrm{B}}T(\mathbf{r})}\right]^{-\kappa-1}\varepsilon. \quad (3)$$

The flux given in Eq.(3) is expanded in terms of the particle energy (logarithms), i.e.,

$$\log j = \left\{\log n + (\kappa-\tfrac{1}{2})\cdot\log(k_{\mathrm{B}}T) - \tfrac{1}{2}\log[(2\pi^{3})m] + \log\left[(\kappa-\tfrac{3}{2})^{\kappa-\tfrac{1}{2}}\frac{\Gamma(\kappa+1)}{\Gamma(\kappa-\tfrac{1}{2})}\right]\right\} - \kappa\cdot\log\varepsilon \quad (4)$$

$$+ O[(\varepsilon_{\mathrm{b}}/\varepsilon)^{\tfrac{1}{2}}] + O[(\kappa-\tfrac{3}{2})k_{\mathrm{B}}T/\varepsilon],$$

indicating that the energy-flux spectra are linearized (on a log-log scale), under the high energy approximation, which is manifested by the conditions $(\kappa-\tfrac{3}{2})k_{\mathrm{B}}T \ll \varepsilon$ and $\varepsilon_{\mathrm{b}} \ll \varepsilon$. Then, we find

$$j = j_{*}\cdot\varepsilon^{-\kappa} \text{ or } \log j = \log j_{*} - \kappa\cdot\log\varepsilon. \quad (5)$$

Namely, the spectral index coincides with the kappa $\kappa$, while the intercept $\log j_{*}$ includes both the notions of temperature and density, i.e.,

$$\log j_{*} = \log n + (\kappa-\tfrac{1}{2})\cdot\log(k_{\mathrm{B}}T) - \tfrac{1}{2}\log[(2\pi^{3})m] + \log\left[(\kappa-\tfrac{3}{2})^{\kappa-\tfrac{1}{2}}\frac{\Gamma(\kappa+1)}{\Gamma(\kappa-\tfrac{1}{2})}\right]. \quad (6)$$

Finally, we write Eq.(6) as

$$\log J_{*} - \log f_{\kappa} = \log n + (\kappa-\tfrac{1}{2})\cdot\log(k_{\mathrm{B}}T), \quad (7)$$

where we set the modified parameter of flux-logarithm intercept, $J_{*}$, and the function of kappa $f_{\kappa}$:

$$J_{*} \equiv j_{*}\cdot\sqrt{(2\pi^{3})\cdot m}, \quad f_{\kappa} \equiv (\kappa-\tfrac{3}{2})^{\kappa-\tfrac{1}{2}}\Gamma(\kappa+1)/\Gamma(\kappa-\tfrac{1}{2}). \quad (8)$$

Next, we show the physical meaning of the flux-logarithm intercept in polytropes. A polytropic process is described by a power-law relation between temperature and density along a streamline of the flow:

$$T(\mathbf{r}) \propto n(\mathbf{r})^{\gamma-1}, \text{ or } n(\mathbf{r}) \propto T(\mathbf{r})^{\nu}, \quad (9)$$

where the polytropic index $\gamma$, and its auxiliary index $\nu$, are given by:

$$\nu = 1/(\gamma-1), \quad \gamma = 1+1/\nu. \quad (10)$$

This relationship implies that

$$\Pi \equiv n(\mathbf{r})\cdot[k_{\mathrm{B}}T(\mathbf{r})]^{-\nu} = const., \quad (11)$$

is an invariant thermodynamic quantity along the polytropic plasma streamline. This is reduced to the thermal pressure in the case of isobaric processes, that is, polytropic processes corresponding to index $v=-1$ or $\gamma=0$. Therefore, the polytropic invariant $\Pi$ generalizes the concept of the invariant thermal pressure $P = nk_BT$ along a streamline for the special polytropes of isobaric processes.

There is a relationship between polytropic index and kappa, $v = -\kappa + \frac{1}{2}$; (see Eq.(A10) in Appendix A, and more details in: Livadiotis 2019a). Hence, the $\Pi$ invariant is kappa dependent

$$\log \Pi = \log[n(r)] + (\kappa - \tfrac{1}{2}) \cdot \log[k_BT(r)]. \tag{12}$$

The expression of the flux in terms of energy, shown in Eq.(5), and especially, the intercept $\log j_*$, shown in Eq.(6), includes the temperature and density with the same form as in the polytropic invariant, shown in Eq.(12). In fact, $j_*$ can be written in terms of $\Pi$ and $\kappa$ alone, i.e.,

$$\log \Pi = \log j_* - \tfrac{1}{2}\log(2\pi^3 m) - \log[(\kappa - \tfrac{3}{2})^{\kappa - \tfrac{1}{2}}\Gamma(\kappa+1)/\Gamma(\kappa - \tfrac{1}{2})], \text{ or } \log \Pi = \log J_* - \log f_\kappa \tag{13}$$

The linear fitting of the energy-flux spectrum of each pixel with Eq.(5) estimates the parameters $\log j_*$ and $\kappa$, from which we determine $J_*$ and $f_\kappa$ via Eq.(8), and the polytropic invariant $\Pi$ via Eq.(13). Therefore, from the $(\varepsilon, j)$-spectrum of each sky-map pixel, we derive a pair of $(\kappa, \Pi)$ values. Then, combining any two pixels, we can solve for the smoothed values of temperature and density.

We note that the observed ENA flux integrates the ENA emissions over the LOS in a sky-map pixel direction $\Omega$ and over the IHS thickness. This leads to the radial average values of the thermodynamic observables (density, temperature, kappa, or their functions, such as the polytropic invariant). Then, the density, temperature, and kappa, involved in the formulation of the linearized energy-flux spectra constitute the radial averages along the radial direction of a LOS through the IHS (Appendix B).

## 3. DATA AND METHOD
### 3.1. Data

This study uses the following referencing indices for the skymaps. The whole sky-map, spanning latitudes $0 \leq \vartheta \leq 180^0$ and longitudes $0 \leq \varphi \leq 360^0$, is binned into 1800 ($6^0 \times 6^0$) pixels, labeled with indices *lat*: 0, 1, …, 29 and *long*: 0, 1, …, 59, so that, a direction at $\Omega(\vartheta, \varphi)$ corresponds to indices *lat*=int($\vartheta$/6) and *long*=int($\varphi$/6) (where int($x$) denotes the integer part of $x$). Note: the map is oriented in such a way so that the upstream direction or "Nose" is located at the center of the map.

The sky-map of ENA observations is separated into the Ribbon and GDF emissions (e.g., see the original papers investigating the Ribbon sources: McComas et al. 2009b, 2010; annual review papers: McComas et al. 2012, 2014, 2017; separation of Ribbon from GDF emissions: Schwadron et al. 2011; 2014; Swaczyna et al. 2022; separation of Ribbon from GDF emissions due to thermodynamic aspects: Livadiotis & McComas 2012). In this paper, we focus on the pristine GDF observations, not by using GDF data separated from the Ribbon, but by using only those sky-map directions with purely GDF emissions – that is, those directions where the Ribbon is not present.

McComas et al. (2020) provided the full solar cycle of IBEX ENA observations and Zirnstein et al. (2021a) subsequently transformed these spectra into the proton plasma frame of the IHS, for the years 2009-2016, using modeled flow profiles derived from a 3D heliosphere simulation to account for the Compton-Getting effect including radial and transverse flows. The values of the proton flux are deduced by assuming a specific value for the IHS width $\Delta r$=30AU. In reality, the width varies over the sky, but this variability does not affect the analysis and results of this study (see Appendix C).

The proton flux is deduced from the ENA flux, according to the model used by Zirnstein et al. (2021a). The proton flux errors are propagated from the observational ENA flux errors. Furthermore, the thermodynamic variables and their errors are derived from the proton flux spectrum and its errors, i.e., the proton flux and its error in each of the IBEX-Hi energy bands. The method uses two fitting processes for the derivation of the temperature and density. When a fitting is involved, two types of uncertainties characterize the results. The total statistical error (not including any systematic errors) of the optimal values of the fitting parameters is composed of two independent errors: (i) the fitting error, which comes from the nonzero square of the residuals, and (ii) the propagation error, which comes from the propagation of the uncertainties characterizing the temperature values. The total error comes the summation of the respective variances (Appendix D). Therefore, the derivation of the temperature and density is based on two types of linear fits: the linear fit of the energy-flux spectrum of each sky-map pixel leading to the values of kappa, $\kappa$, and invariant $\Pi$, as well as to their fitting and propagation errors. Then, combining a number of neighboring pixels, we produce the values of temperature and density, smoothed over this selection of neighboring pixels, as well as their fitting and propagation errors.

### 3.2. Description of the method

It is now straightforward to determine the values of temperature and density of each pixel. We deal with two unknowns involved in a single equation (underdetermined solution), i.e.,

$$\log n + (\kappa - \tfrac{1}{2}) \cdot \log(k_B T) = \log J_* - \log f_\kappa = \log \Pi . \quad (14)$$

This problem is overcome by applying Eq.(14) to two neighboring sky-map directions. In particular, the smoothing analysis (Appendix D) considers common (i.e., smoothed) values of temperature and density for two neighboring pixels; then, Eq.(14) can be determined for two pixels: (1) the pixel at the examined direction $\Omega = (\vartheta, \varphi)$, and (2) a neighboring pixel at the direction $\Omega' = (\vartheta', \varphi')$; hence, we have a linear system of two equations, where $\log(k_B T)$ and $\log n$ are the two common unknowns:

$$\log n + [\kappa(\Omega) - \tfrac{1}{2}] \cdot \log(k_B T) = \log \Pi(\Omega) \ , \ \log n + [\kappa(\Omega') - \tfrac{1}{2}] \cdot \log(k_B T) = \log \Pi(\Omega') \ . \quad (15)$$

These two equations are combined to solve in terms of the density and temperature,

$$\log[k_B T(\Omega', \Omega)] = \frac{\log[\Pi(\Omega')/\Pi(\Omega)]}{\kappa(\Omega') - \kappa(\Omega)} \ , \ \log n(\Omega', \Omega) = \frac{\log \Pi(\Omega')/[\kappa(\Omega') - \tfrac{1}{2}] - \log \Pi(\Omega)/[\kappa(\Omega) - \tfrac{1}{2}]}{1/[\kappa(\Omega') - \tfrac{1}{2}] - 1/[\kappa(\Omega) - \tfrac{1}{2}]} \quad (16)$$

Therefore, the temperature and density of each pixel is smoothed for two neighboring pixels, namely, by solving Eq.(15) for the two pixels. Then, we repeat for all the eight neighboring pixels, producing eight values of $(T,n)$ (Figure 1). As we observe in Eq.(16), the derivation $(T,n)$ requires a significant variability between the values of intercept and spectral index or $(\log \Pi, \kappa)$ characterizing the flux spectra of neighboring pixels. Finally, we perform statistical analysis for averaging the derived set of eight values of $(T,n)$, under certain statistical confidence (Appendix D).

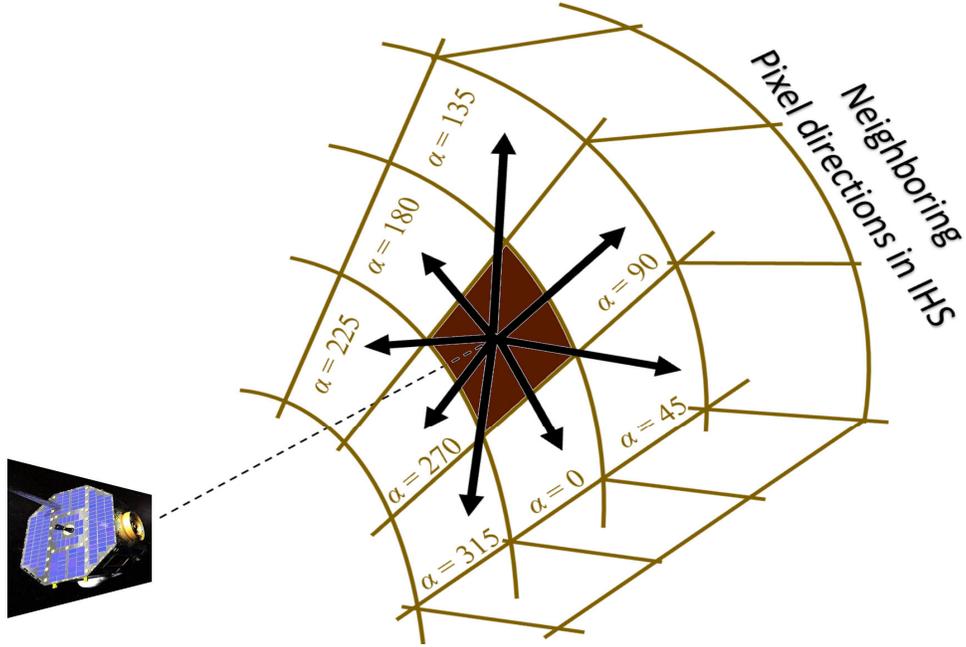

**Figure 1.** Each sky-map pixel is surrounded by eight closest neighboring pixels; (here the examined central pixel is colored brown). Each combination of the central pixel with one neighboring pixel is used to estimate a pair of temperature and density values, ($T,n$). Then, we perform a statistical analysis for averaging the derived set of eight values of ($T,n$). The procedure is repeated for each sky-map pixel.

The smoothness of temperature and density makes sense if they do not vary too much from pixel to pixel; but how much variation is allowed for the smoothing to be accepted? Typically, the goodness of fitting a constant (that is, of averaging) is accepted when the fitting uncertainty (caused by the variation of the values) is significantly small, that is, smaller than the (average) errors of the individual fit points. The exact statistical tools that characterize the goodness of the fitting that averages the eight estimated values of ($T,n$) are the $p$-value or the reduced chi-square (as explained in Appendix D).

Figure 2(a)(b) shows an example of good fit, as measured by condition of high $p$-value. The panels show the temperatures of the eight pixels surrounding the central pixel located at the direction $\Omega$ with longitude and latitude indices (long=13, lat=9), determined for the 2016 observations; (the temperatures are presented with respect to the projected angle $\alpha$). Panel 2(c) shows the fraction of all 1270 GDF sky-map pixels that correspond to a significant $p$-value, i.e., larger than a certain confidence level for the averaging fit to be accepted; the fraction is plotted with respect to this confidence level. There are ~52% pixels with $p$-value > 0.05 and ~75% pixels with $p$-value > 0.01 (both the confidence levels of 0.05 and 0.01 are commonly used). Panel 2(d) shows the estimated error of the logarithm of temperature, which measures the normalized temperature error $\delta \log T \sim 0.434 \cdot \delta T / T$, plotted with respect to the $p$-value of the fitting. The red points indicate the average value of $\delta \log T$ within binned $p$-values with width $\Delta(p\text{-value}) = 0.1$; the blue line shows the trend of these points, where we observe a slow decrease of $\delta \log T$ with increasing the $p$-value. (For more details on the statistical analyses, and the involved conditions of $p$-value and reduced chi-square, see: Appendix D).

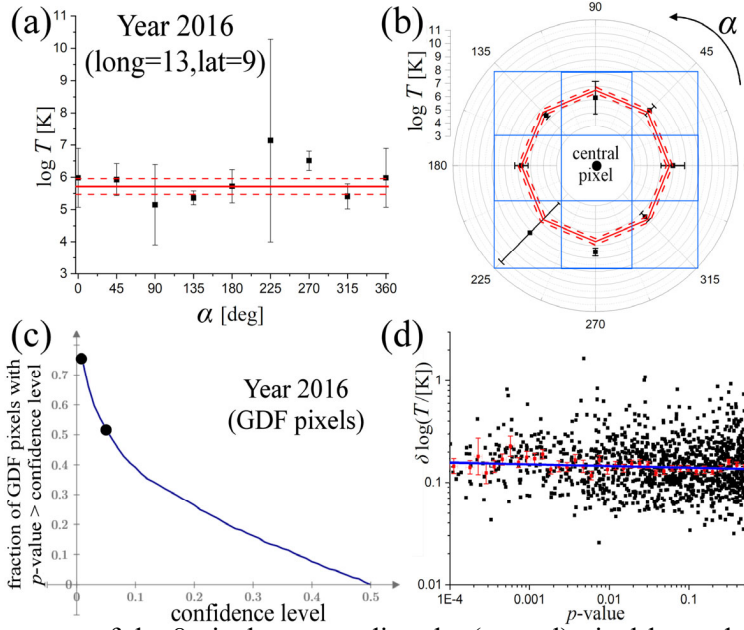

**Figure 2.** (a) Temperatures of the 8 pixels surrounding the (central) pixel located at (long=13,lat=9) for year 2016, presented with respect to the projected angle $\alpha$ (with $\alpha = 360$ deg simply repeating the case of $\alpha = 0$ deg). This is typical of what is observed in the skymaps of GDF. (b) Similar to (a), but plotted in polar coordinates. (c) Fraction of all GDF sky-map pixels of year 2016 with $p$-values above some confidence level; (the two dots indicate the commonly used levels of 0.05 and 0.01, corresponding to fractions ~52% and ~75%). (d) Estimated error of the (logarithm of) temperature, plotted on a log-log scale for all the sky-map pixels of the year 2016, with respect to the $p$-value. The red points indicate the average value of $\delta \log T$ within binned $p$-values with bin width $\Delta(p\text{-value}) = 0.1$; the blue line shows the trend of these points, that is, $\log(\delta \log T) \approx -0.859 \pm 0.013 - (0.0173 \pm 0.0067) \cdot \log(p\text{-value})$.

Figure 3 shows another example, where we consider the condition of small reduced chi-square (together with small relative errors), i.e., $\chi^2_{\text{red}} < 3$ and $\delta \log T \leq 0.5$, instead of the condition of high $p$-value, i.e., $p > 0.01$. Panel (a) plots the temperatures from the eight pixels surrounding the central pixel at LOS vector $\Omega$: (long=13, lat=12), for year 2012; panel (b) magnifies the temperature axis of panel (a). We observe that the error of the mean is quite smaller than the data errors, i.e., $\delta c^*_{\text{st}} \ll \delta c^*_{\text{pr}}$. The $p$-value is small (< 0.01), but the sum of residuals or standard deviation is also small ($\chi^2_{\text{red}} \ll 1$), and thus the chi-square can be conditionally accepted, examining the derived temperature to have sufficiently small error. Indeed, this is $\delta \log T \sim 0.2$, smaller than the upper threshold of half order of magnitude. Panels (c) and (d) plot the cumulative percentage of all the GDF pixels of 2012 up to a certain $p$-value or $\chi^2_{\text{red}}$, respectively. We observe that (a) ~58.7% of the pixels correspond to $p$-value > 0.05, but (b) ~96.8% of them to $\chi^2_{\text{red}} \leq 3$ (and $\delta \log T \leq 0.5$).

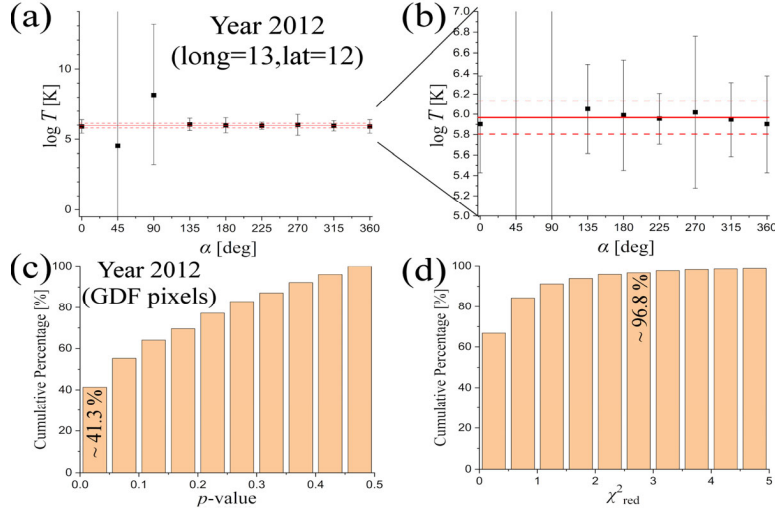

**Figure 3.** (a) Temperatures of the 8 pixels surrounding the (central) pixel located at the LOS direction $\Omega$ with longitude and latitude indices (long=13,lat=12), for the year 2012, presented with respect to the projected angle $\alpha$ (similar to Figure 1(a)). (b) Magnification of (a), showing the small residuals compared to the respective errors of each point. The plotted cumulative percentages of (c) $p$-values and (d) reduced chi-square $\chi^2_{\rm red}$ show that 41.3% of the GDF pixels have $p$-value < 0.05 (or 58.7% of the GDF pixels with $p$-value $\geq$ 0.05); (note that there is ~80% of the GDF pixels with $p$-value $\geq$ 0.01 – not shown in the graph); however, there is a larger percentage of 96.8% of GDF pixels with $\chi^2_{\rm red} \leq 3$ (and $\delta \log T \leq 0.5$).

### 3.3. Verification of the method

There are four basic considerations in the method of this work: (i) the velocities/energies of the protons in the IHS are described by kappa distributions, (ii) the energy-flux spectra are produced from high-energy protons (above the core of the proton population), (iii) smoothing neighboring pixels implies that there is no (statistically significant) variation of temperature and density values; and (iv) there is no significant impact of systematic uncertainties of the observations used in this work. Next, we discuss these considerations, providing satisfactory verifications.

*3.3.1. Verification of the kappa function.*

The effects of a kappa distributed source in the heliosheath on the global heliosphere and expected ENA fluxes at 1 AU was studied theoretically even before IBEX data was available (e.g., Heerikhuisen et al. 2008). As shown in Eq.(5), the power-law behavior of energy-flux spectra can be explained by the formulation of kappa distributions and their power-law tails; this leads to the linearity of these high-energy spectra on log-log scales. Certainly, the kappa function is not the only model that extends to a power-law tail; (e.g., Baliukin et al. 2020;2022). Nevertheless, the "signature" of kappa distributions is the unique factor $f_\kappa$, which appears in the flux vs. energy relationship (Eqs.(5-8)), having its origin in the normalization of kappa distributions and their high-energy approximation (Appendix A). Indeed, the logarithm of flux scales linearly with the logarithm of energy, according to Eq.(5), while the intercept $\log j_*$ depends on kappa so that the difference $\log j_* - \log f_\kappa$ becomes linearly related to kappa,

$$\log j_{*_i} - \log f_{\kappa_i} = \log n_i + (\kappa_i - \tfrac{1}{2}) \cdot \log(k_{\rm B} T_i) + const., \qquad (17)$$

for $i$ indicating each GDF pixel of each observation year. In the original paper of Livadiotis et al. (2011), the above linearity was suggested to characterize the whole sky map. This may not be true for pixels in ribbon directions owing to the superposition of ENA fluxes from physically different sources at different heliocentric distances. However, for a single population such as the GDF, if the temperature and density are less variant than the kappa, then we may write $\log n_i \sim \langle \log n \rangle$ and $\log(k_B T_i) \sim \langle \log(k_B T) \rangle$, or

$$\log j_{*_i} - \log f_{\kappa_i} = \kappa_i \cdot \langle \log(k_B T) \rangle + const.' , \qquad (18)$$

The linearity of this function with $\kappa$ is a characteristic of kappa distributions. Figure 4(a) plots the values of $\log \Pi_i$ or $\log j_{*_i} - \log f_{\kappa_i}$ against $\kappa_i$ for all the GDF pixels in 2012. We observe two main populations of temperature, $\langle \log(k_B T / [K]) \rangle = 6.049 \pm 0.012$ and $6.147 \pm 0.018$, which agree with the most frequent temperature in the histogram plotted in Figure 4(b).

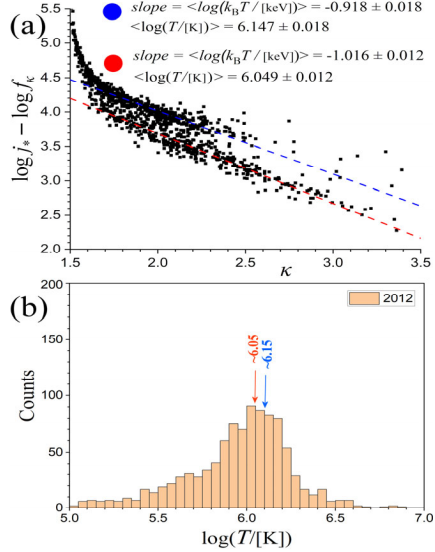

**Figure 4.** The linearity of $\log \Pi$, i.e., the function $\log j_* - \log f_\kappa = const. + (\kappa - 1/2) \cdot \log(k_B T)$, with kappa $\kappa$ is a characteristic property of the kappa function. This linearity is shown in (a), plotting the values of $\log j_{*_i} - \log f_{\kappa_i}$ against $\kappa_i$ for all the GDF pixels of 2012. The two main linear streams correspond to temperatures, $\langle \log(k_B T / [K]) \rangle = 6.049 \pm 0.012$ (red) and $6.147 \pm 0.018$ (blue), agreeing with the most frequent temperature value shown in the histogram plotted in (b).

*3.3.2. Verification of the high-energy proton population.*

The condition for having a proton plasma population in the high-energy region that describes the power-law tail is $\varepsilon \gg (\kappa - 3/2)k_B T$ and $\varepsilon \gg \varepsilon_b \equiv \frac{1}{2} m u_b^2$, which is fulfilled for the vast majority of GDF sky-map pixels. For values of kappa $1.6 < \kappa < 3$ and temperatures $5 < \log(T/[K]) < 6.5$, with most frequent values $\kappa \sim 1.8$ and $\log(T/[K]) \sim 6$, we obtain that $(\kappa - \frac{3}{2})k_B T$ ranges from $\sim 0.8$ eV to $\sim 0.8$ keV with most frequent value $\sim 20$ eV. Also, $u_b$ ranges from 100-200 km/s, i.e., $\varepsilon_b \sim 50$-200 eV. Thus, both $(\kappa - \frac{3}{2})k_B T$ and $\varepsilon_b$ are significantly smaller than the energies of IBEX-Hi channels that range from a half to a few keV.

The energy-flux spectra of kappa distributed protons are reduced to a power-law at the high-energy limit, where the spectral index equals the kappa value characterizing the distribution. Then, we would have expected to have the same observations for the kappa and spectral indices. However, this is not true, because in order for the observed spectra to be connected with a kappa distribution, two requirements are necessary: the power-law behavior is the first condition, from which we determine the spectral index and the intercept; the second condition is the linear fitting between the values of spectral index, $\kappa_i$, and the modified intercept, or $\log \Pi$ values, $\log j_{*_i} - \log f_{\kappa_i}$ (described in 3.3.1). This last condition, fulfilled under a satisfactory statistical confidence, ensures that (i) the spectrum expression is described by a kappa function, (ii) the kappa value coincides with the spectral index, and (iii) the value of log Π can be used to determine the temperature and density.

Figure 5 shows the histograms of the kappa and spectral indices of all the GDF pixels and the annual sky-maps for the examined years (2009 – 2016). The kappa values have narrower range, because they are restricted by the requirements of (i) the limiting condition of $\kappa > 3/2$, and (ii) the linearity between kappa and log Π invariant values. The observed most frequent values (modes) of kappa and spectral indices are ~1.8 and ~2.0, respectively.

It has to be noted that if the high energy limit of kappa distributions is not being reached, then, the spectrum deviates from a power-law (i.e., it is slightly curved, compared to the linear behavior plotted on a log-log scale); being concave, even slightly, the spectral index of the observed power-law won't reflect the kappa value of the distribution but rather some less steep value. Nevertheless, the statistical analysis (Appendix D) requires a good fitting characterizing the linearity of spectra (on log-log scale), and thus, the possibility of having a curved spectrum is largely eliminated.

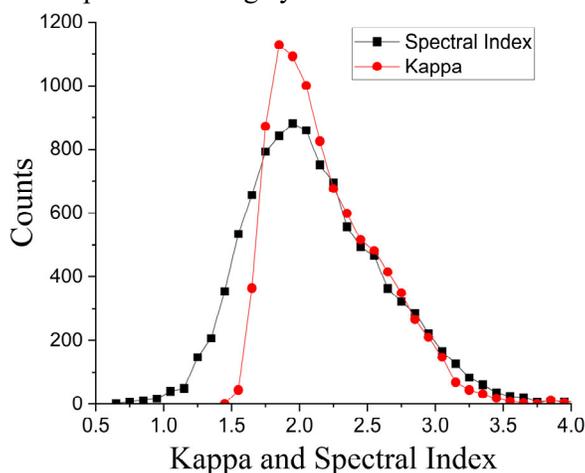

**Figure 5.** Overall histogram of the kappa and spectral indices for the examined observation years 2009 – 2016; the observed modes are ~1.8 and ~2.0, respectively.

*3.3.3. Verification of the smoothed temperature and density values of neighboring pixels.*

We find that the majority of all the GDF pixels are characterized by a statistically confident smoothing. For all the examined years 2009–2016, there are 1270 pixels corresponding to purely GDF emissions; the percentage of GDF pixels with acceptable constant fit to the neighboring temperatures is: (a) 63%±8% with respect to the *p*-value condition (*p*-value ≥ 0.01), and (b) 87%± 2% with respect to the chi-square condition ($\chi^2_{\mathrm{red}} \leq 3$ and $\delta \log T \leq 0.5$); for further details on these conditions, see: Appendix D).

In Appendix D.2.3 we calculate the overall distinction of the pixels according to the goodness of the fitting that provides the smoothed temperature and densities of neighboring pixels. Figure 6 plots the annual percentages of the distinguishing categories, and for each of the observational years between 2009 and 2016, while the inset pie-chart shows their all-years mean values.

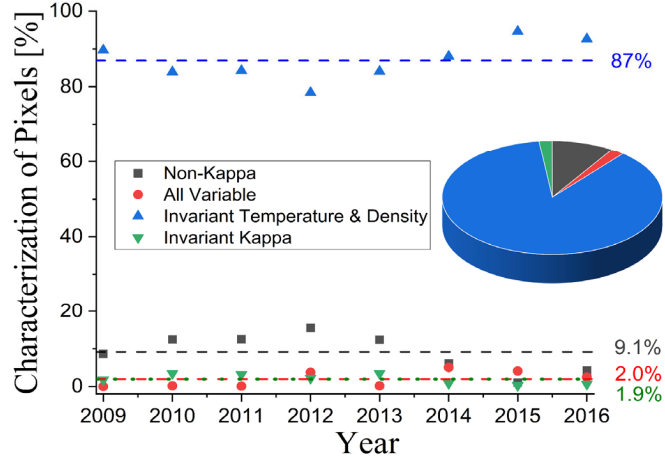

**Figure 6.** Annual percentage of GDF sky-map pixels, corresponding to: (i) non-kappa behavior, that is, estimated kappa is less than 1.5 (grey); (ii) all variable (i.e., no statistical constant value of $T$, $n$, and $\kappa$ for neighboring pixels) (red); (iii) invariant $T$, $n$, and variable $\kappa$, (i.e., statistical constant value for neighboring pixels) (blue); and (iv) invariant $\kappa$ (statistically constant value of $\kappa$, hence, unknown value of $T$ and $n$) (green). Pie-chart shows the all-years mean values.

*3.3.4. Verification of non-significant impact of small systematic uncertainties.*

The statistical errors of spectral flux have already been considered in the derivation of the temperature and the other thermodynamic variables (Appendix D). The question we address here is in regards to the impact of systematic errors. IBEX-Hi Electrostatic Analyzer (ESA) -2 has the highest systematic uncertainty (including background noise and instrumental uncertainties) among the five IBEX-Hi energy bands used in this paper. However, the consideration of one-point systematic error, especially when this point corresponds to ESA-2 that has already been characterized by a large statistical error and thus less statistical weight in the involved analyses, would not lead to a significant deviation of the estimated thermodynamic variables. Nevertheless, a systematic error of the spectral index would strongly affect the determined temperature (and related thermodynamic variables).

Spectral index systematic errors require systematic flux uncertainties suitably distributed among the energy bands. The flux systematic errors, noted by $\Delta j$, must have affected in different portions the flux of each energy band; then, the spectral flux of each energy band, $j(\varepsilon_2)$, $j(\varepsilon_3)$, $j(\varepsilon_4)$, $j(\varepsilon_5)$, $j(\varepsilon_6)$, becomes after the impact of systematic errors, $j(\varepsilon_2)+\Delta j_2$, $j(\varepsilon_3)+\Delta j_3$, $j(\varepsilon_4)+\Delta j_4$, $j(\varepsilon_5)+\Delta j_5$, $j(\varepsilon_6)+\Delta j_6$. We may choose a gradual increasing in the flux so that the slope systematically increases with increasing energy; i.e., $\Delta j_i = \sigma \cdot c_i \cdot \delta j_i$, with $c_i$ increasing with $i$, $\delta j_i$ is the respective statistical uncertainty of the flux of the $i^{\text{th}}$ energy band (where the index $i$ increases with energy), and $\sigma$ stands for the highest absolute ratio of the systematic per statistical errors among energy bands. For instance, we may have $c_2 := -1$, $c_3 := -1/2$, $c_4 := 0$, $c_5 := +1/2$, $c_6 := +1$, where $\sigma := |\Delta j_2/\delta j_2| := |\Delta j_6/\delta j_6|$ (that is, corresponding to ESA-2 and ESA-6).

Now we can address the question of how large should the ratio of systematic-to-statistical errors be in order for the results to be significantly affected by the systematic errors. In Figure 7, we consider systematic errors distributed along energy bands ($c_i$ as mentioned; the maximum error ratio is noted by $\Delta j/\delta j$). Figure 7(a) plots the flux spectrum for the pixel at (long=0,lat=1) from 2012 observations, either with no systematic errors (black) and with the systematic error of $\Delta j/\delta j = 0.6$ (red), where fluxes become $j(\varepsilon_2)-0.6\delta j$, $j(\varepsilon_3)-0.5\cdot 0.6\delta j$, $j(\varepsilon_4)$, $j(\varepsilon_5)+0.5\cdot 0.6\delta j$, $j(\varepsilon_6)+0.6\delta j$. Next, given a milder systematic error with $\Delta j/\delta j = 0.4$, we examine the impact on the estimated temperature; in particular, we derive the temperature systematic error $\Delta \log T$, normalized by its (original) statistical error $\delta \log T_0$, for each sky-map pixel; then, we plot the distribution of the values of $\Delta \log T/\delta \log T_0$ in Figure 7(b). We measure the half width at half maximum (HWHM); for instance, panel (b) indicates HWHM ~ 1.18. Figure 7(c) plots HWHM with respect to the maximum error ratio $\Delta j/\delta j$; the linear fit shows that $\Delta \log T / \delta \log T_0$ is about 3 times the value of $\Delta j/\delta j$.

In general, the systematic errors are unknown and not accounted in this analysis. However, we were able to find that the estimated temperature values may vary significantly with the impact of systematic errors of spectral fluxes, only if all the following occur: (i) the systematic errors are more than ~1/3 of the respective statistical errors, (ii) the systematic errors are organized in an increasing (or decreasing) manner along the energy band (e.g., as modeled by $c_i$), and (iii) the systematic errors affect a large number of the sky-map pixels.

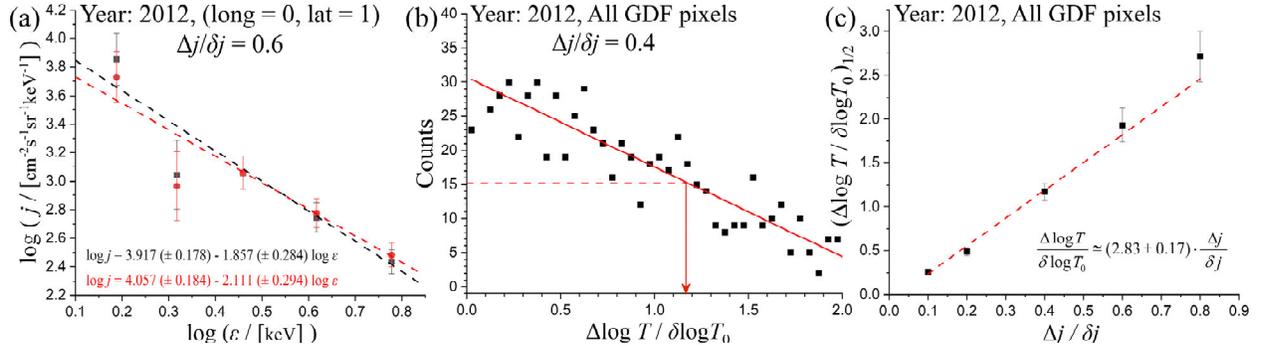

**Figure 7.** (a) Flux spectrum for the pixel at (long=0,lat=1) from 2012 observations, plotted with no systematic errors (black) and with the systematic error of $\Delta j/\delta j = 0.6$ (red). (b) Histogram of the temperature systematic error normalized by its (original) statistical error, $\Delta \log T/\delta \log T_0$, for each sky-map pixel during 2012 observations; the estimated HWHM is ~ 1.18. (c) Plots of HWHM with respect to the maximum error ratio $\Delta j/\delta j$; the linear fit shows that $\Delta \log T / \delta \log T_0$ is about 3 times the value of $\Delta j/\delta j$.

## 4. RESULTS
### 4.1. Annual sky-maps

Figure 8 shows the resulting annual sky-maps of the logarithms of the mean flux (flux averaged over the IBEX-Hi energy passbands ESA 2-6), temperature, density, and kappa, derived for all the years of GDF ENA observations, spanning from 2009 to 2016 (eight annual maps).

We note a number of general characteristics: (1) Small fluctuations of temperature, density, and kappa values, for the majority of the sky-map directions in the IHS; (2) Lower temperature and kappa, higher density, near the nose; opposite behavior near the flanks and tail; (3) Correlations: positive between temperature and kappa, and negative between temperature and density; (4) Temperature sky-maps are shifted to their lowest values near 2012 and to their highest near 2015.

SKY MAPS

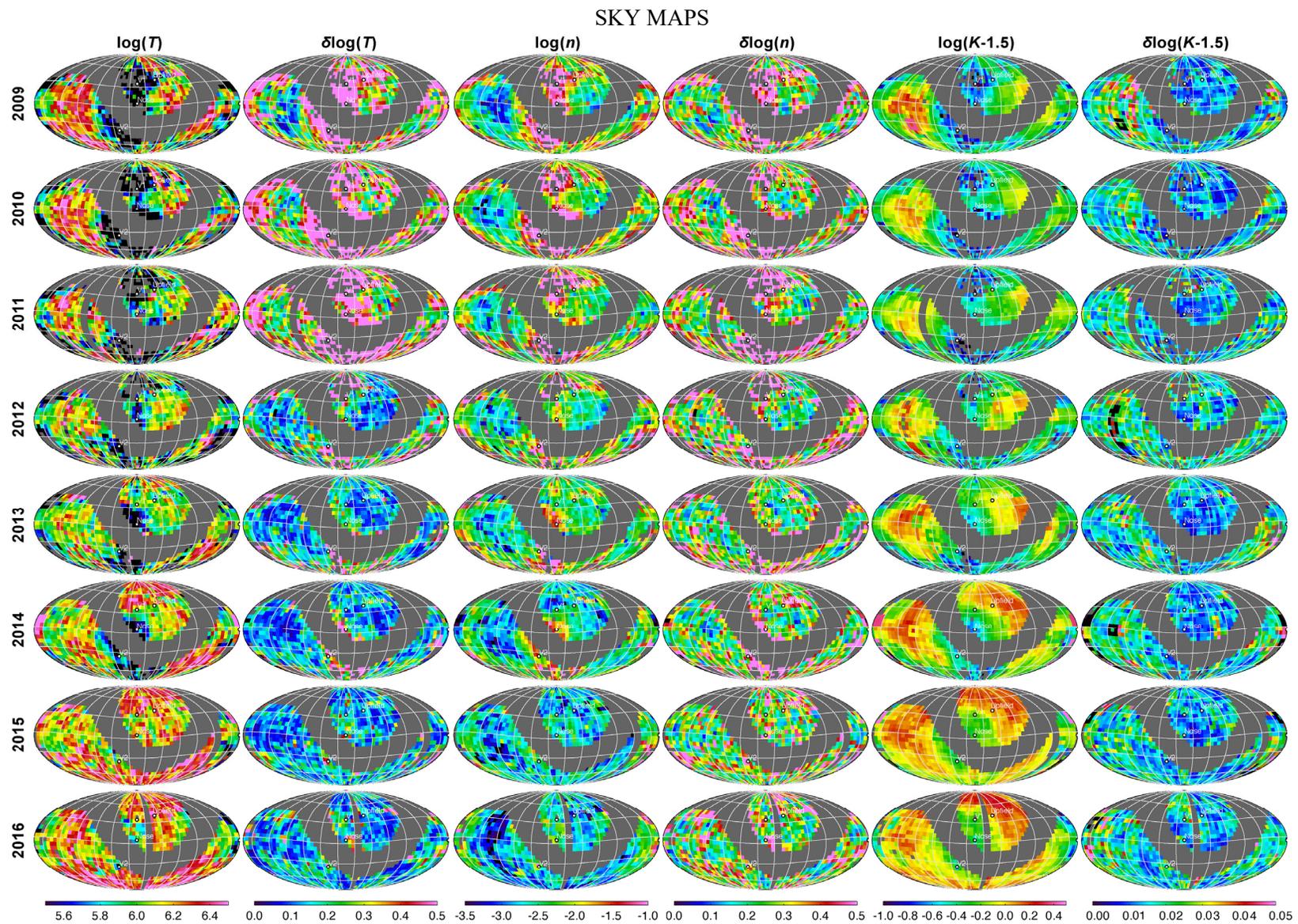

**Figure 8.** Annual sky-maps of the temperature, density, and kappa, plotted on log scales together with their uncertainties for the years 2009 - 2016.

## 4.2. Annual histograms

Figure 9 plots the 1-dimensional histograms (unnormalized occurrence) of the temperature logarithm values, corresponding to all the sky-map GDF pixels for each year. We observe that the histograms are (i) characterized by small asymmetry and dominant modes (in this case, the mode, instead of the mean, provides a characteristic value of the temperature; (ii) the histograms are shifted to the lowest temperatures in 2012 and to the highest temperatures in 2015; and (iii) the two histograms around the minimum temperatures, that is, over 2012 and 2013, are relatively similar, and the same holds for the two histograms around the maximum temperatures, that is, over 2015 and 2016; this might indicate stabilization near an extremum.

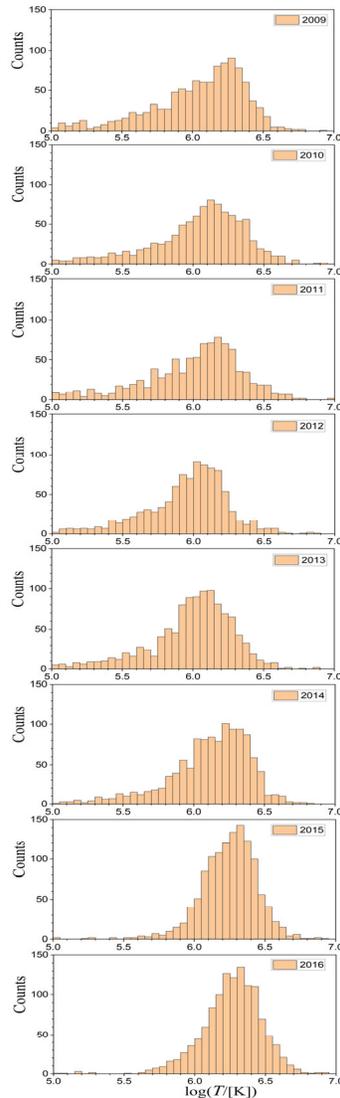

**Figure 9.** Histograms (unnormalized occurrence) of the temperature logarithm values, taken from all the GDF sky-map pixels and derived separately for each year of ENA observations.

## 4.3. Evolution of the modes of histograms

We fit a Gaussian distribution to the values of the logarithm of temperature and calculate the modes of the distributions and their uncertainty. Figure 10 plots the modes of the annual histograms of the temperature (shown in Figure 9), density, and kappa, for all the sky-map GDF pixels of the ENA observation years from 2009 to 2016. A positively correlated trend is observed for the mean flux, temperature, and kappa, and a respective negative trend is observed for the density, though there is a deviation from the linear trend in the early observation years between 2009 - 2011. The highest (negative) correlation appears between the annual values of the temperature and density. This is related to the polytropic relationship of sub-isothermal polytropic index, $\gamma < 1$, examined in more detail in Section 5).

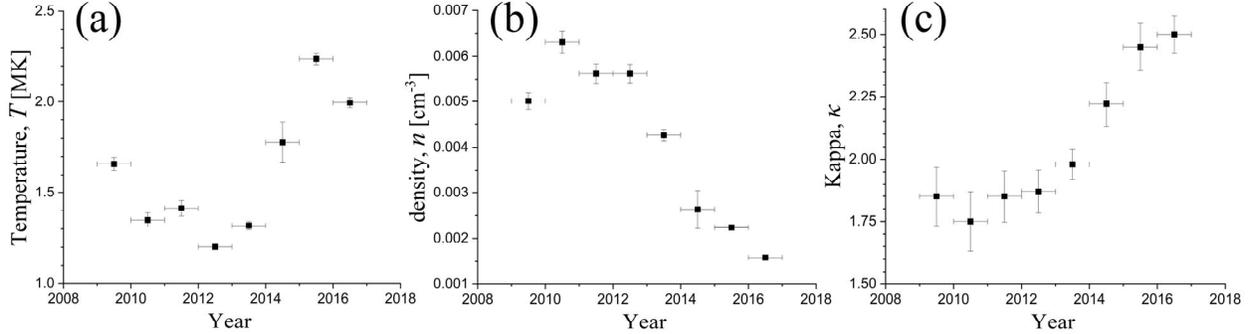

**Figure 10.** Annual values and standard deviation of the modes of the (a) temperature, (b) density, and (c) kappa, plotted for each year of ENA observations.

In Figure 10(a) a single Gaussian distribution was used to fit the histograms. Instead, we may consider two statistically significant populations, and fit the histogram with a superposition of two Gaussians. We find that during the years of ENA observations, the coolest and hottest years of the IHS was in 2012 and 2015, respectively. Figure 11 plots the superposition of two Gaussians that fits the annual data of the logarithms of temperature for the years (a) 2012 and (b) 2015. The temperatures for both the populations and their weighted mean is shown in (c).

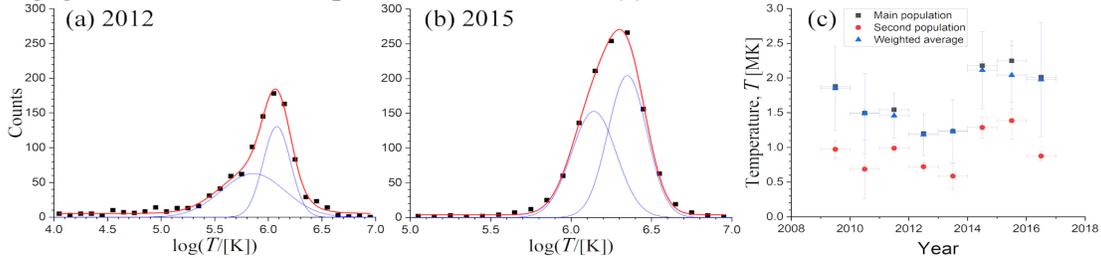

**Figure 11.** A superposition (red) of two Gaussians (blue) fits the histograms for (a) 2012 and (b) 2015. (c) Derived temperatures for both populations and their weighted mean for all the observation years.

### 4.4. Characteristics of the sky-maps and histograms
*4.4.1. Variation of thermodynamic variables along streamlines*

In the sky-maps of Figure 8, we observe lower temperature and kappa values, as well as higher density values near the nose, while the opposite behavior is observed toward the flanks and tail.

The thermodynamic parameters (including kappa) may vary along streamlines. Ideal polytropic flows allow the variation of temperature and density, but not kappa. More complicated phenomenological polytropic flows may be needed for correctly describing the observed variation of thermodynamic variables (e.g., Livadiotis 2016). In addition, physical mechanisms other than polytropes may also be involved. For instance, the thermodynamic mixing of the proton plasma population with solar wind protons in higher latitudes increases the particle entropy shifting the system into stationary states of higher kappa values; on the other hand, the mixing with newly formed pickup protons decreases the particle entropy because of the highly ordered motion of these ions, shifting the system into stationary states of lower kappa values (for details, see: Livadiotis & McComas 2010; 2011a; 2011b); depending on which mixing mechanism prevails we may observe the kappa values to increase or decrease along the streamlines from the nose to flanks and tail. Finally, other mechanisms varying the thermodynamic observables on streamlines may also exist, such as the interaction of charge exchange cross-section with temperature and their thermodynamic parameters (e.g., Zirnstein & McComas 2015; Chalov 2018).

*4.4.2. Differences between spectral indices and kappa values*

As described in Section 3 and Figure 5, the differences between the kappa and spectral indices are caused by (i) the limiting condition of $\kappa > 3/2$, and (ii) the linearity between kappa and log $\Pi$ invariant values. Figure 12 compares the spectral indices and kappa values.

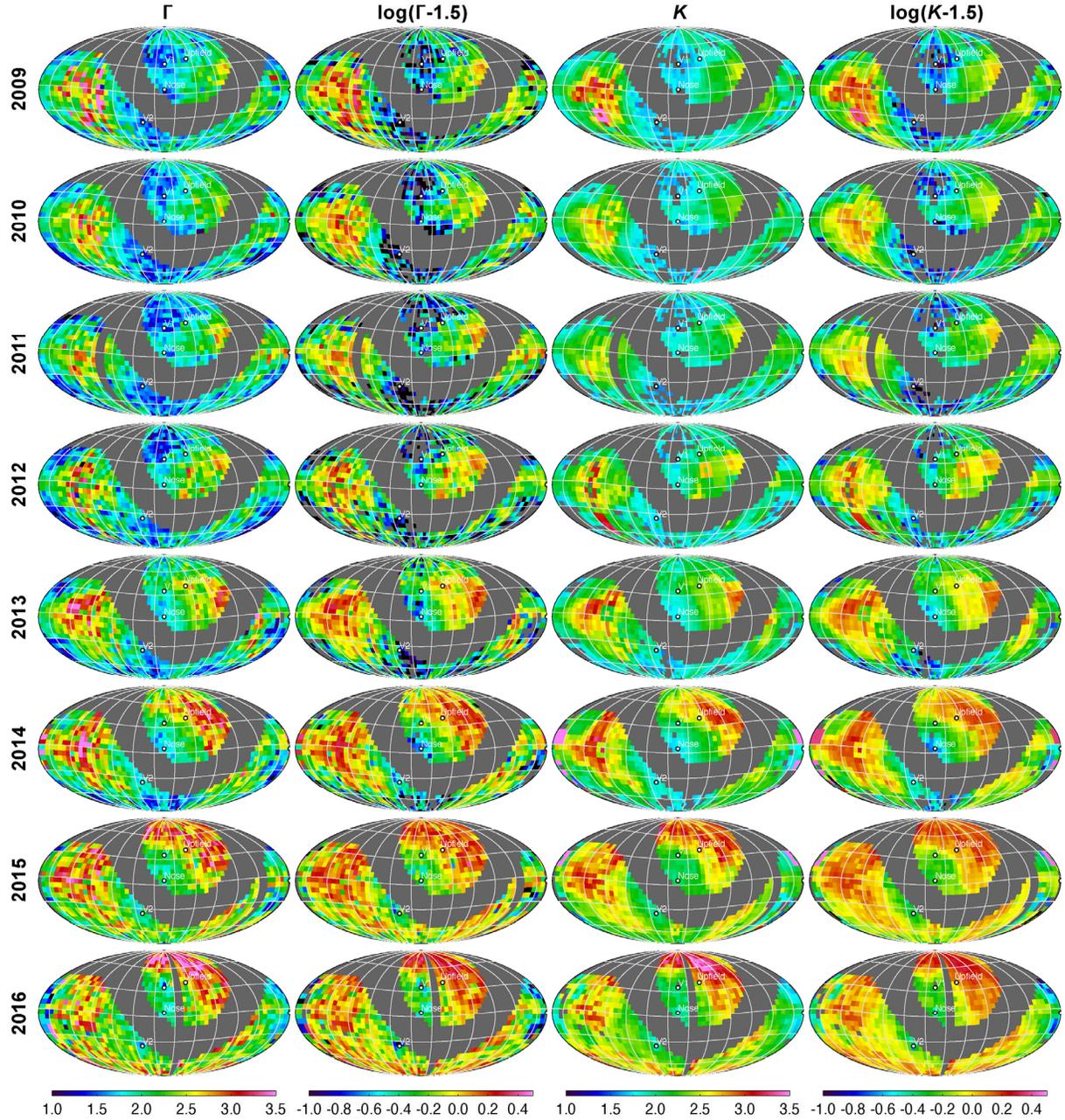

**Figure 12.** Annual sky-maps of the spectral indices ($\Gamma$) and kappa ($\kappa$), plotted also on log scales together with their uncertainties for the years 2009 - 2016.

*4.4.3. Correlations:*

(1) Temperature is negatively correlated with density. This behavior is closely related to the polytropic thermodynamic process that characterize the proton plasma in the IHS through its flow along the streamlines from the nose to the tail. In particular, the polytropic index that characterizes these processes is sub-isothermal, $\gamma < 1$, which leads to the observed negative correlation between $T$ and $n$, since the polytropic behavior involves $T \propto n^{\gamma-1}$, i.e., $T \propto 1/n^{1-\gamma}$. This result was first shown for the thermodynamic parameters of the proton in the IHS over the 2009 ENA IBEX observations (Livadiotis & McComas 2012; 2013b; Livadiotis 2016), while it is aligned to the underlying theory that connects the polytropic index with turbulent heating (Fahr & Chashei 2002; Livadiotis 2019b; 2021).

(2) Temperature is positively correlated with kappa. A local $\kappa$–$T$ relationship is frequently observed in space plasmas. For example, the positively correlated kappa and temperature values in the magnetospheres of Earth (Ogasawara et al. 2013; 2015), Jupiter (Collier & Hamilton 1995), and Saturn (Dialynas et al. 2009; 2018). Such a relationship has already been detected in the IHS, using the first IBEX observations over 2009 (Livadiotis & McComas 2012). The observed relationships between kappa and temperature do not violate their global independence, which is a fundamental property coming from thermodynamic laws. In particular, it was shown that the most generalized formulation aligned with the concept of temperature is the kappa distributions, where kappa and temperature are independent parameters, assigned with their own kinetic and thermodynamic definitions (Abe 2001; 2002; Toral 2003; Livadiotis 2018b; Livadiotis & McComas 2021). The observed relationships between the temperature and kappa are local properties of the plasma and they do not imply a universal equation; they are connected to the mechanisms responsible for generating kappa distributions in space plasmas. Some examples are: superstatistics (Beck & Cohen 2003; Schwadron et al. 2010; Hanel et al. 2011); effect of shock waves (Zank et al. 2006); weak turbulence (e.g., Yoon 2014; Bian et al. 2014); effect of pickup ions (Livadiotis & McComas 2011a;b); pump acceleration mechanism (e.g., Fisk & Gloeckler 2014); charge-exchange (Heerikhuisen et al. 2015); polytropic behavior (Livadiotis 2018d; 2019a); Fokker–Planck equation (Shizgal 2018); common processes characteristic of space plasmas, such as the Debye shielding and magnetic coupling, have an important role in the generation of kappa distributions in plasmas (Livadiotis et al. 2018).

Figure 13(a) displays the occurrence frequency of the temperature logarithm values, filtered with small error, $\delta \log T < 0.5$, over the 2009-2016 years of observation.

Given the sampling distribution of $\log(T/[\text{K}])$ values, it is not surprising that the (unnormalized) 2D-histogram of $\log(n/[\text{cm}^{-3}])$ vs. $\log(T/[\text{K}])$ in Figure 13($b_1$) also shows a dominant global maximum, corresponding to the mode observed in (a). We, therefore, normalize the 2D-histograms to investigate the actual relationship between the thermodynamic parameters, and here, $\log(T/[\text{K}])$ vs $\log(n/[\text{cm}^{-3}])$. Figure 13($c_1$) shows the 2D-histogram normalized by the 1D-histogram of $\log(T/[\text{K}])$ plotted in (a), which clearly demonstrates that the temperature and density are negatively correlated. The plot of the weighted mean and the standard error for each $\log(T/[\text{K}])$ – bin, for all bins, is shown in Figure 13($d_1$). Similarly, in Figure 13($b_2$) and 13($c_2$), we plot the unnormalized and normalized 2D-histograms of $\log(T/[\text{K}])$ with $\log(\kappa\text{-}3/2)$; panel 13(d2) plots the weighted means and standard errors; the corresponding normalized histogram reveals that the temperature and kappa are positively correlated.

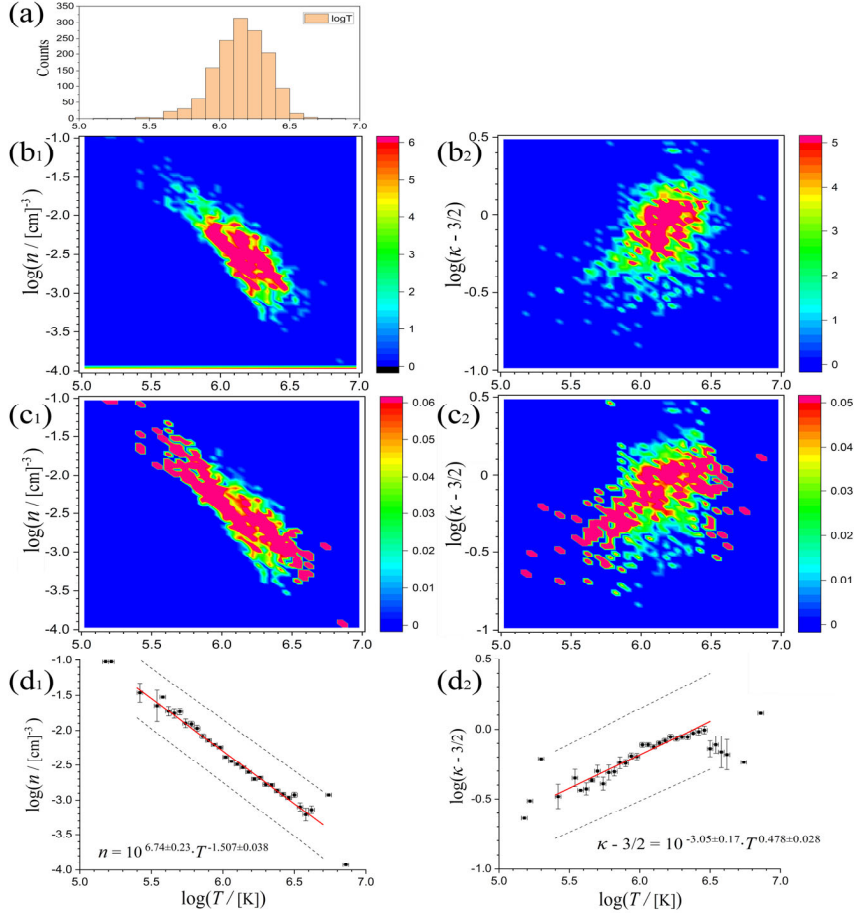

**Figure 13.** (a) Histogram of logarithms of temperature values, log($T$/[K]), plotted for all the examined ENA observation years (2009-2016). ($b_1$) 2D-histogram of the all-year values of log($n$/[cm$^{-3}$]) vs. log($T$/[K]). ($c_1$) 2D-distribution from panel ($b_1$) normalized by the histogram of log($T$/[K]) values, plotted in (a). ($d_1$) Plot of weighted mean and the standard error for each log($T$/[K]) – bin. ($b_2$-$d_2$): Same as ($b_1$-$d_1$) but with examining variables log($T$/[K]) and log($\kappa$-3/2). Color-bars measure the smoothed 2D occurrence frequency.

## 5. DERIVATION OF ASSOCIATED THERMODYNAMIC PARAMETERS
### 5.1. Polytropic Index

The derivation of the polytropix index $\gamma$ involves finding the exponent in the polytropic power-laws $T \propto n^{\gamma-1}$ or $p \propto n^{\gamma}$. Equivalently, we may derive the secondary polytropic index $\nu$, the exponent in the polytropic power-laws $n \propto T^{\nu}$ or $p \propto T^{\nu+1}$, where $\nu = 1/(\gamma-1)$, $\gamma = 1 + 1/\nu$ (Section 2, Appendix A).

There are various methods for deriving a polytropic index, which can be separated in two main categories, (i) solving for the polytropic index from plasma dynamics equations, and (ii) addressing the polytropic relationship between thermodynamic variables, e.g., $T \propto n^{\gamma-1}$ or $p \propto n^{\gamma}$. Some examples of the former method (i) are the Rankin-Huguenot condition for energy for finding the polytropic index around a shock (e.g, Tatrallyay et al. 1984; Winterhalter et al. 1984) or the sound wave speed formulation for finding the polytropic index of the medium (e.g., Gardiner et al. 1998).

The latter method (ii) is more common and refers to two main types of data analyses that examine: (1) the timeseries of the evolution of the thermodynamic variables, and/or (2) the macroscopic behavior of the thermodynamic variables in the whole dataset. In type (1), the timeseries of $\{\log n_i, \log T_i\}_{i=1}^{N}$ is known and the polytropic index is derived from (a) fitting linearly the sequential values of $\{\log n_i, \log T_i\}_{i=1}^{N}$ with the slope being equal to $\gamma$-1 (e.g., Nicolaou et al. 2014), or of $\{\log n_i, \log p_i\}_{i=1}^{N}$ with the slope equal to $\gamma$ (e.g., Baumjohann & Paschman 1989); (b) solving the polytropic index in terms of the sequential values, $\gamma_i = 1 + \log(T_{i+1}/T_i)/\log(n_{i+1}/n_i)$ or $\gamma_i = \log(p_{i+1}/p_i)/\log(n_{i+1}/n_i)$, followed by a running average of the derived indices $\{\gamma_i\}_{i=1}^{N}$ (e.g., Livadiotis & Desai 2016). In type (2), the polytropic index is derived from a global linear fitting (on a log-log scale) of all the given values of the evolved thermodynamic parameters, which may not be in a sequential timeseries (e.g., Nicolaou et al. 2020), i.e., $\{\log n_i, \log T_i\}_{i=1}^{N_{all}}$ or $\{\log n_i, \log p_i\}_{i=1}^{N_{all}}$.

The macroscopic analysis is simpler than the more accurate analyses of sequential values; however, it can be similarly accurate, if there is: (i) a single statistical population of polytropic indices, and (ii) absence of points with extreme deviations. In particular, the condition (i) is tested by examining the statistical confidence of one-population fitting (typically using the *p*-value and the reduced chi-square for measuring the goodness of the fitting, as explained is Section 3 and Appendix D). Moreover, the condition (ii) is tested by examining the distribution of the fluctuating slopes: Each value of the fluctuating slopes is derived from macroscopic linear fitting using all the data points except one point, which is excluded from the fitting, and is different for each derivation of the fluctuating slope; then, we construct a set of $N_{all}$ fluctuating slopes, whose distribution must (1) have mean value equal to the fitted slope and (2) be extended within the fitted optimal slope error (of the overall fitting).

As an example, Figure 14(a) shows the derivation of the polytropic index for the year 2010. The fitting is unweighted, because we observed it maximizes the correlation between temperature and density values and minimizes the error of the polytropic index. In the case of an unweighted fitting, the minimum chi-square can be calculated as follows: The fitting or statistical type of error, $\sigma_{st}$, is generally given by the sum of the square residuals divided by the degrees-of-freedom (DOF) $M = 1103$; we find $\sigma_{st}^2 \cong 0.067$. The most frequent error of the data is $\sigma^2 \cong (\delta \log T)^2 + (\gamma-1)^2 (\delta \log n)^2 \cong 0.06$, where $\delta \log T \cong 0.21$, $\delta \log n \cong 0.16$, and $\gamma - 1 \cong -0.81$. Hence, $\chi_{red}^2 \cong \sigma_{st}^2/\sigma^2 \cong 1.1$ and $p-\text{value} \cong 0.011$, while the correlation coefficient is 0.95, leading to an acceptable fitting goodness. Figure 14(b) plots the histogram of all the fluctuating slopes. We observe that (1) the mean of the histogram is the same with the slope of the linear fitting in (a); and (2) the histogram is extended within the standard error of the linear fitting in (a). Figure 14(c) plots the derived polytropic index for all the years of ENA observations, from 2009 to 2016.

The results show that the polytropic index is smaller than the isothermal, $\gamma < 1$, as predicted by the theory (Livadiotis 2021), corresponding to a polytropic proton plasma residing in a sub-isothermal state. Occasionally, the polytropic index falls near zero (especially during the solar minimum), corresponding to a polytropic isobaric state. The IHS is characterized as sub-isothermal tending to be isobaric.

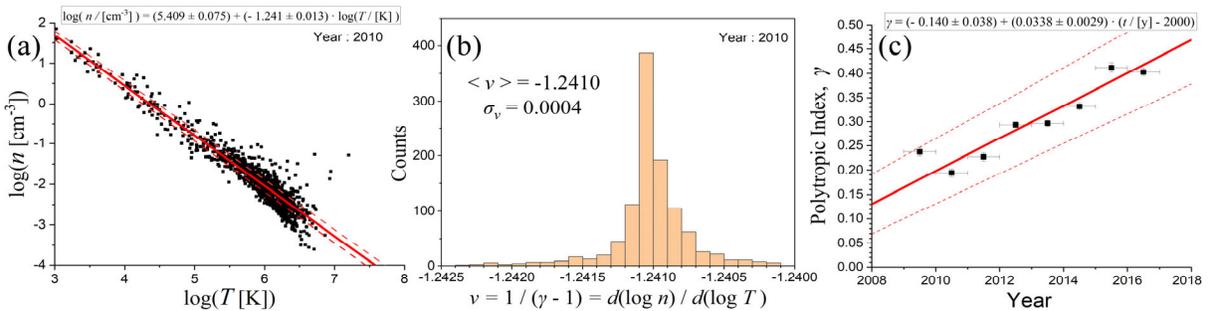

**Figure 14.** (a) Secondary polytropic index, $\nu \equiv 1/(\gamma - 1)$, derived macroscopically from the slope of temperature-density diagram (on log-log scales). (b) Histogram of the (secondary) polytropic indices (distribution of fluctuating slopes – see text). (c) Evolution of the polytropic index $\gamma = 1 + 1/\nu$.

The theory behind the connection of kappa distributions and the polytropic behavior leads to the relationship between the kappa and the polytropic index, where the total potential energy that sums gravitational, electric, and magnetic components may be involved (Livadiotis 2018d; 2019a).

Figure 15 plots the polytropic index, (a) $v$, and (b) $\gamma$, with respect to the annual modes of the kappa for the eight years of ENA observations examined by this study. The theory predicts that the secondary polytropic index $v$ is negatively correlated with kappa, thus the standard polytropic index $\gamma$ is positively correlated; a linear model is fitted to both the graphs. These relationships can be used to understand the underlying potential energy among the particles (e.g., Livadiotis 2018d; Nikolaou & Livadiotis 2019).

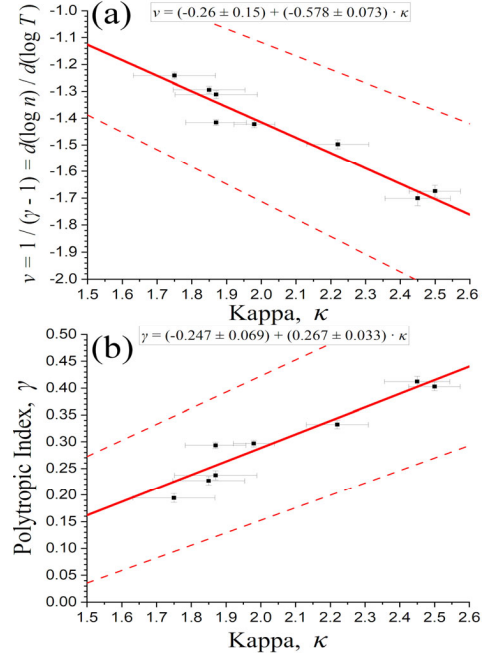

**Figure 15.** Annual polytropic indices plotted with respect to the kappa values for the years of ENA observations between 2009 – 2016. A linear model is fitted to the data.

*5.2. Entropy*

The entropy is explicitly expressed in terms of the density, temperature, and kappa. In particular, the entropy for kappa distributions is written as $S(n,T,\kappa) = \kappa \cdot [1 - e^{-\frac{1}{\kappa} \cdot S_\infty(n,T)}]$, where $S_\infty = S_\infty(n,T)$ denotes the classical, density and temperature dependent, entropy. (For details on the formulation, see: Appendix E; Livadiotis 2018b; Livadiotis & McComas 2021; see also Adhikari et al. 2020).

Given their annual values, shown in Figure 10, we calculate the respective annual values of entropy, depicted in Figure 16(a), while Figure 16(b) shows their plot with respect to the polytropic index. The entropy annual values may be impacted by the solar cycle and the bi-modal properties of solar wind. We also observe that the entropy increases as polytropic index increases towards the isothermal state of $\gamma = 1$. The isothermal state corresponds to the Maxwellian distribution ($\kappa \rightarrow \infty$) (Livadiotis & McComas 2011a; Livadiotis 2019a). These findings are aligned to the theory of kappa distributions, because the polytropic index $\gamma$ is positively correlated to the kappa values, as shown in Figure 15(b); as $\gamma$ increases, kappa increases, i.e., the kappa distribution shifts to Maxwell-Boltzmann distribution (Livadiotis et al. 2018).

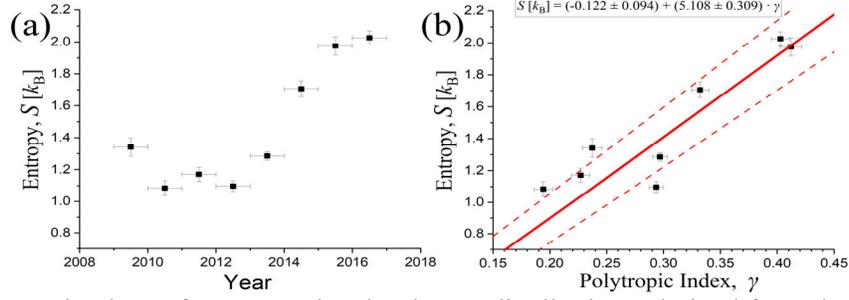

**Figure 16.** (a) Annual values of entropy related to kappa distributions, derived from the annual values of the density, temperature, and kappa, which they were plotted in Figure 10. (b) Annual values of entropy shown in (a), plotted with respect to the annual polytropic indices, which were plotted in Figure 14(c).

### 5.3. Temperature percentage rate

For each GDF sky-map pixel, we calculate the annual percentage rate change of the temperature, $d\log T / dt \sim \Delta\log T / \Delta t$ (with $\Delta t = 1y$), and then construct the histogram and its mode for the values within one year of ENA observations. In Figure 17(a) we plot these annual values of $d\log T / dt$ for each year of observation; (the variation between two sequential years is assigned at the year boundary between the two years involved). In Figure 17(b) we re-plot the respective annual values of temperature, for comparison. We observe a general decreasing of temperature in the years 2009-2012 and mostly an increasing of temperature in the years 2012-2016. The rate is near zero between 2012 and 2013 and between 2015 and 2016, that is, similar to the time period where the temperature has its lowest and highest annual values (extrema), respectively.

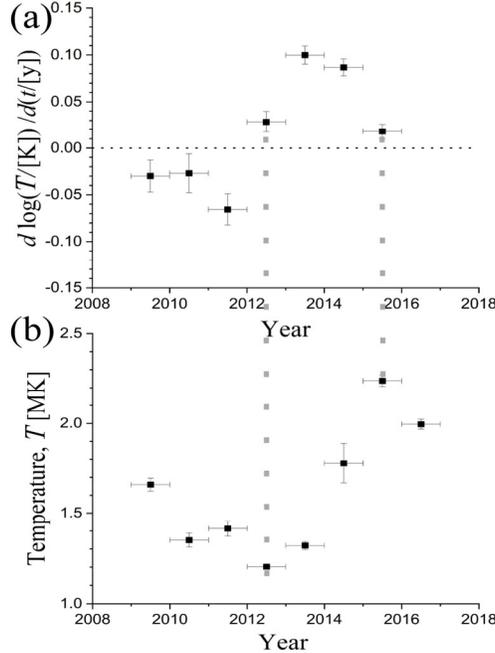

**Figure 17.** (a) Annual percentage rate of temperature (mean and standard error). (b) Annual temperature values (also shown in Figure 10a).

### 6. CONCLUSIONS

We derived the sky maps of temperature and other thermodynamic parameters of the proton plasma in the IHS, such as, density, kappa, polytropic index, and entropy, for all the directions characterized by purely GDF ENA emissions over the years 2009 – 2016 of IBEX observations within the solar cycle 24.

We exploited the theory of kappa distributions and their connection with polytropes to (i) define a new polytropic quantity Π that remains invariant along plasma flow streamlines, despite variations of the temperature and density, (ii) parameterize the proton flux in terms of Π and kappa $\kappa$, (iii) use the pair of values of Π and $\kappa$ of each pixel to express the values of temperature and density.

In the method, we determined the temperature and density along the LOS in the IHS for each sky-map pixel, performing three statistical analyses: (i) fitting of the proton flux spectra, leading to the values of $\kappa$ and Π of each pixel; (ii) smoothing of the temperature and density values, by combining the values of ($\kappa$, Π) of a pixel with those values of each of its eight neighboring pixels, leading to a set of eight smoothed values of temperature and density – one for each neighboring pixel; and (iii) averaging of the derived eight smoothed values. The method was verified via three statistical analyses that examine: (i) the kappa formulation involved in energy-flux spectra; (ii) the high-energy limit of the formulation of kappa distributions; and (iii) the variation of kappa values for neighboring pixels.

We summarize the results of this study: (i) sky-maps of temperature, density, and kappa; (ii) temperature annual histograms; (iii) evolution of the histogram modes of the mapped parameters; (iv) polytropic indices; (v) entropy; (iv) temperature rate;

With regards to thermodynamic characteristics of the IHS, we observed the following:
(1) *Temperature sky-maps and histograms shifted* to their lowest values near 2012 and to their highest near 2015.
(2) *Temperature negatively correlated with density*. This behavior is due to the polytropic thermodynamic process characterizing the proton plasma in the IHS through its flow from the nose toward the tail. In particular, the polytropic index that characterizes these processes is sub-isothermal, $\gamma < 1$, which leads to the observed negative correlation (Livadiotis & McComas 2012; 2013b).
(3) *Temperature positively correlated with kappa*. This is a local plasma property rather than a universal one between thermodynamic parameters. In general, these trends are caused by the various mechanisms responsible for generating kappa distributions in space plasmas (Livadiotis et al. 2018).
(4) *Polytropic index is sub-isothermal*, i.e., smaller than the isothermal value, $\gamma < 1$, aligned with earlier derivations (Fahr & Chashei 2002; Elliot et al. 2019; Livadiotis 2021), corresponding to a polytropic proton plasma residing in a sub-isothermal state. During solar minimum, the polytropic index falls near zero, corresponding to a polytropic isobaric state. Therefore, the IHS is characterized as sub-isothermal tending to be isobaric during solar minimum.
(5) *Linear relationship between annual kappa and polytropic indices*: The theory predicts specific correlations among these thermodynamic parameters, which can be used to determine the underlying particle potential energy (e.g., Livadiotis 2018d; Nikolaou & Livadiotis 2019).
(6) *Entropy increases with increasing the polytropic index towards the isothermal state* ($\gamma = 1$), corresponding to the Maxwellian distribution ($\kappa \to \infty$) (Livadiotis & McComas 2010; 2011a).

Acknowledgements. This work was carried out under the interstellar Boundary Explorer (IBEX) mission (80NSSC20K0719), which is a part of NASA's Explorer Program.

**Appendix A. Formulation of the kappa distribution function**

The kappa distribution of velocities ***u*** or energy $\varepsilon$ (Binsack 1966; Olbert 1968; Vasyliũnas 1968) is

$$P(\varepsilon;T,\kappa) = \frac{(\kappa-\tfrac{3}{2})^{-\tfrac{3}{2}}}{B(\tfrac{3}{2},\kappa-\tfrac{1}{2})} \cdot (k_B T)^{-\tfrac{3}{2}} \cdot \left[1 + \frac{1}{\kappa-\tfrac{3}{2}} \cdot \frac{\varepsilon}{k_B T}\right]^{-\kappa-1} \varepsilon^{\tfrac{1}{2}}, \tag{A1}$$

with normalization $1 = \int_0^\infty P(\varepsilon;T,\kappa)d\varepsilon$, where $\varepsilon$ expresses the particle kinetic energy $K(\boldsymbol{u}) = \tfrac{1}{2}m|\boldsymbol{u}-\boldsymbol{u}_\mathrm{b}|^2$ (in the frame of the flow, where $\boldsymbol{u}_\mathrm{b}$ is the plasma flow velocity). This formulation generalizes the Maxwell-Boltzmann distribution and applies when plasma particles reside in stationary states out of the classical thermal equilibrium (Livadiotis 2018b) in the absence of any interactions (Livadiotis 2015c).

In the presence of particle interactions, the one-particle phase-space distribution function $f(\boldsymbol{r},\boldsymbol{u})$ is described by the kappa function of a Hamiltonian $H$, while the one-particle Hamiltonian, $H(\boldsymbol{r},\boldsymbol{u}) = \Phi(\boldsymbol{r}) + K(\boldsymbol{u})$, is written as the sum of a position-depended potential energy $\Phi(\boldsymbol{r})$, and the velocity-depended kinetic energy $K(\boldsymbol{u}) = \tfrac{1}{2}m|\boldsymbol{u}-\boldsymbol{u}_\mathrm{b}|^2$ (e.g., Jurac et al. 2002; Livadiotis 2015c). Then,

$$f(\boldsymbol{r},\boldsymbol{u};T,\kappa) = n_\infty \cdot \pi^{-\tfrac{3}{2}} \cdot (\kappa-\tfrac{3}{2})^{-\tfrac{3}{2}} \frac{\Gamma(\kappa+1)}{\Gamma(\kappa-\tfrac{1}{2})} \cdot (\tfrac{2}{m}k_\mathrm{B}T)^{-\tfrac{3}{2}} \cdot \left[1 + \frac{1}{\kappa-\tfrac{3}{2}} \cdot \frac{\Phi(\boldsymbol{r}) + \tfrac{1}{2}m|\boldsymbol{u}-\boldsymbol{u}_\mathrm{b}|^2}{k_\mathrm{B}T}\right]^{-\kappa-1}, \quad (A2)$$

with the normalization $1 = \int_{-\infty}^{+\infty} P(\boldsymbol{r},\boldsymbol{u};T,\kappa)d\boldsymbol{r}d\boldsymbol{u}$ or $N = \int_{-\infty}^{+\infty} f(\boldsymbol{r},\boldsymbol{u};T,\kappa)d\boldsymbol{r}d\boldsymbol{u}$, where $N$ is the number of particles; $k_\mathrm{B}$ is the Boltzmann's constant; symbols $d\boldsymbol{r}$ and $d\boldsymbol{u}$ abbreviate the infinitesimal volume in positional and velocity space, respectively; the probability distribution, $P = f/N$, may be used instead of the distribution function $f$; and, $d$ is the total DOF, defined by the ensembled average of the Hamiltonian, $d \equiv 2\langle H(\boldsymbol{r},\boldsymbol{u})\rangle/(k_\mathrm{B}T)$, or $d = d_\mathrm{K} + d_\Phi$, that is, equal to the sum of the kinetic DOF $d_\mathrm{K} \equiv 2\langle K(\boldsymbol{r},\boldsymbol{u})\rangle/(k_\mathrm{B}T)$ (here $d_\mathrm{K} = 3$) and the potential DOF, $d_\Phi \equiv 2\langle \Phi(\boldsymbol{r})\rangle/(k_\mathrm{B}T)$. (For more details, see, e.g., Livadiotis 2019a).

In the absence of a potential energy, the one-particle Hamiltonian describes the particle's kinetic energy, $H(\boldsymbol{u}) = K(\boldsymbol{u}) = \tfrac{1}{2}m|\boldsymbol{u}-\boldsymbol{u}_\mathrm{b}|^2$, and the distribution function is formulated by the standard function:

$$f(\boldsymbol{u};n_\infty,T,\kappa) = n_\infty \cdot \pi^{-\tfrac{3}{2}} \cdot (\kappa-\tfrac{3}{2})^{-\tfrac{3}{2}} \cdot \frac{\Gamma(\kappa+1)}{\Gamma(\kappa-\tfrac{1}{2})} \cdot (\tfrac{2}{m}k_\mathrm{B}T)^{-\tfrac{3}{2}} \cdot \left[1 + \frac{1}{\kappa-\tfrac{3}{2}} \cdot \frac{\tfrac{1}{2}m|\boldsymbol{u}-\boldsymbol{u}_\mathrm{b}|^2}{k_\mathrm{B}T}\right]^{-\kappa-1}, \quad (A3)$$

with the normalization: $1 = \int_{-\infty}^{+\infty} P(\boldsymbol{u};T,\kappa)d\boldsymbol{u}$, $N = \int_{-\infty}^{+\infty} f(\boldsymbol{u};n_\infty,T,\kappa)d\boldsymbol{u}$. The distribution is complicated in the presence of a potential energy; but it can be reduced to the exact form as if there were no potential energy. We have:

$$f(\boldsymbol{r},\boldsymbol{u};n_\infty,T,\kappa) = n_\infty \cdot \left[1 + \frac{1}{\kappa-\tfrac{d}{2}} \cdot \frac{\Phi(\boldsymbol{r})}{k_\mathrm{B}T}\right]^{-\kappa+\tfrac{1}{2}} \pi^{-\tfrac{3}{2}} \cdot (\kappa-\tfrac{d}{2})^{-\tfrac{3}{2}} \frac{\Gamma(\kappa+1)}{\Gamma(\kappa-\tfrac{1}{2})} \cdot \left(\tfrac{2}{m}k_\mathrm{B}T \cdot \frac{\kappa-\tfrac{d}{2}}{\kappa-\tfrac{3}{2}}\right)^{-\tfrac{3}{2}}$$

$$\times \left\{1 + \frac{1}{\kappa-\tfrac{3}{2}} \cdot \frac{\tfrac{1}{2}m|\boldsymbol{u}-\boldsymbol{u}_\mathrm{b}|^2}{k_\mathrm{B}T \cdot \frac{\kappa-\tfrac{d}{2}}{\kappa-\tfrac{3}{2}} \cdot \left[1 + \frac{1}{\kappa-\tfrac{d}{2}} \cdot \frac{\Phi(\boldsymbol{r})}{k_\mathrm{B}T}\right]}\right\}^{-\kappa-1}, \text{ or} \quad (A4)$$

$$f[\boldsymbol{u};n(\boldsymbol{r}),T(\boldsymbol{r}),\kappa] = \pi^{-\tfrac{3}{2}} \cdot (\kappa-\tfrac{3}{2})^{-\tfrac{3}{2}} \frac{\Gamma(\kappa+1)}{\Gamma(\kappa-\tfrac{1}{2})} \cdot n(\boldsymbol{r}) \cdot [\tfrac{2}{m}k_\mathrm{B}T(\boldsymbol{r})]^{-\tfrac{3}{2}} \cdot \left[1 + \frac{1}{\kappa-\tfrac{3}{2}} \cdot \frac{\tfrac{1}{2}m|\boldsymbol{u}-\boldsymbol{u}_\mathrm{b}|^2}{k_\mathrm{B}T(\boldsymbol{r})}\right]^{-\kappa-1}. \quad (A5)$$

Therefore, the kappa function of a Hamiltonian (A2) has been rewritten as when there is no potential energy. The latter is enclosed in the formulation of the position-depended density and temperature:

$$T(\boldsymbol{r}) = T_\infty \cdot \left[1 + \frac{1}{\kappa-\tfrac{d}{2}} \cdot \frac{\Phi(\boldsymbol{r})}{k_\mathrm{B}T}\right] \text{ and } n(\boldsymbol{r}) = n_\infty \cdot \left[1 + \frac{1}{\kappa-\tfrac{d}{2}} \cdot \frac{\Phi(\boldsymbol{r})}{k_\mathrm{B}T}\right]^{-\kappa+\tfrac{1}{2}}, \quad (A6)$$

where $n_\infty$ and $T_\infty$ are the density and temperature at positions where the potential becomes zero (usually, at infinity), while $T$ provides the globally average temperature,

$$T \equiv \langle T(\boldsymbol{r})\rangle = T_\infty \cdot \frac{\kappa-\tfrac{3}{2}}{\kappa-\tfrac{d}{2}}. \quad (A7)$$

Relations (A6) reveal the connection of kappa distributions with the polytropic behavior. Typically, this is given by the following behavior of temperature and density along a streamline of the plasma flow:

$$T(\boldsymbol{r}) \propto n(\boldsymbol{r})^{\gamma-1} \text{, or } n(\boldsymbol{r}) \propto T(\boldsymbol{r})^{\nu}, \tag{A8}$$

where the polytropic index $\gamma$, and its auxiliary index $\nu$, are given by:

$$\nu = 1/(\gamma-1) \text{ , } \gamma = 1 + 1/\nu. \tag{A9}$$

Comparing Eqs.(A8,A9), we find the relationship between polytropic index and kappa:

$$\nu = -\kappa + \tfrac{1}{2} \text{ , } \gamma = 1 + 1/(-\kappa + \tfrac{1}{2}). \tag{A10}$$

Lastly, we mention the concept of the invariant kappa, $\kappa_0$. The typical notation of kappa is actually depended on the dimensionality of the system, and specifically, on the particle kinetic (=3) and potential (=$d_\Phi$) degrees of freedom,

$$\kappa(d) = \kappa_0 + \tfrac{1}{2}d = \kappa_0 + \tfrac{3}{2} + \tfrac{d_\Phi}{2}. \tag{A11}$$

The notion of the invariant kappa is independent of the dimensionality and thus, it is suitable for characterizing the thermodynamics of the system, and formulating the multi-dimensional distributions (Swaczyna et al. 2019). In the case of the stationary state corresponding to the classical thermal equilibrium, kappa to $\kappa_0 \to \infty$. In the case of the furthest possible stationary state from the classical thermal equilibrium (anti-equilibrium) (Livadiotis & McComas 2013a), kappa tends to $\kappa_0 \to 0$. Each stationary state is described by a kappa distribution of a certain kappa, with the two limits (i) at the classical thermal equilibrium for $\kappa_0 \to \infty$, described by the Maxwell-Boltzmann velocity distributions and zero correlation among particle velocities, and (ii) at anti-equilibrium, where the velocity distribution tends to a simple power-law (Livadiotis & McComas 2010) and the correlation among particle energies is maximized (Abe 1999; Asgarani & Mirza 2007; Livadiotis & McComas 2011a; Livadiotis 2015b; Livadiotis et al. 2021). (For more details on the theory of kappa distributions and its connection to nonextensive statistical mechanics and thermodynamics, see: the books of Livadiotis 2017 and Yoon 2019; the reviews in Livadiotis & McComas 2009; Pierrard & Lazar 2010; Livadiotis & McComas 2013a, Livadiotis 2015a; and the pioneer paper of Treumann 1997; and Milovanov & Zelenyi 2000; Leubner 2002; Livadiotis & McComas 2009. In regards to nonextensive statistical mechanics, see: Tsallis 1988; 2009.)

**Appendix B. Radial averages along a LOS within the IHS**

Here we show that the values of the density, temperature, and kappa, involved in the energy-flux proton spectra, represent their radial averages along the LOS within the IHS (Section 2).

The original data of the observed ENA flux involve the integration of the proton flux along the line-of-sight (LOS) (Gruntman 1992), that is, approximately along the radial direction, within the IHS, i.e.,

$$j_{\text{ENA}} \propto \int_{r \in \text{LOS}} j(\boldsymbol{r}) d\boldsymbol{r} \propto \langle j \rangle \cdot \Delta V. \tag{B1}$$

The ENA flux $j_{\text{ENA}}$ has been converted to the proton (average) flux $\langle j \rangle$, whose power-law behavior at high energies gives $j \propto j_* \propto nT^{\kappa-\tfrac{1}{2}}$ (as shown in Eqs.(5,6)); hence, we have

$$\langle j \rangle \cdot \Delta V = \int_{r \in \text{LOS}} j(\boldsymbol{r}) d\boldsymbol{r} \propto \int_{r \in \text{LOS}} j_*(\boldsymbol{r}) d\boldsymbol{r} \propto \int_{r \in \text{LOS}} n(\boldsymbol{r}) T(\boldsymbol{r})^{\kappa-\tfrac{1}{2}} d\boldsymbol{r}$$

$$= \frac{\int_{r \in \text{LOS}} n(\boldsymbol{r}) d\boldsymbol{r}}{\int_{r \in \text{LOS}} d\boldsymbol{r}} \cdot \frac{\int_{r \in \text{LOS}} n(\boldsymbol{r}) T(\boldsymbol{r})^{\kappa-\tfrac{1}{2}} d\boldsymbol{r}}{\int_{r \in \text{LOS}} n(\boldsymbol{r}) d\boldsymbol{r}} \cdot \int_{r \in \text{LOS}} d\boldsymbol{r} = \langle n \rangle \cdot \langle T \rangle_p^{\kappa-\tfrac{1}{2}} \cdot \Delta V, \tag{B2}$$

$$\text{or, } \langle j \rangle \propto \langle n \rangle \cdot \langle T \rangle_p^{\kappa-\tfrac{1}{2}}, \tag{B3}$$

which includes: (i) the radial average of the density, (ii) the radial power-law average (or $p$-mean, Abramowitz & Stegun 1965) of temperature, and (iii) the IHS volume along the LOS; i.e., respectively,

$$\langle n \rangle = \frac{\int_{r \in \text{LOS}} n(r) dr}{\int_{r \in \text{LOS}} dr}, \quad \langle T \rangle_p = \left[ \frac{\int_{r \in \text{LOS}} n(r) T(r)^p dr}{\int_{r \in \text{LOS}} n(r) dr} \right]^{\frac{1}{p}} \quad \text{with } p = \kappa - \tfrac{1}{2}, \text{ and } \Delta V = \int_{r \in \text{LOS}} dr, \quad (B4)$$

where the $p$-exponent in Eq.(B2) is given by $p = \kappa - 1/2$. Note: The streamlines within the IHS may be characterized by different kappa values; then, along the LOS across the IHS, the kappa should vary with the radial distance, since different distance will cross different streamline. In this case, the concept of $p$-means is generalized to that of $\Phi$-means (e.g., Kolmogorov 1930; Aczel 1948; Czachor & Naudts 2002).

**Appendix C. Dependence on the IHS width of the proton plasma fluxes**

The values of the proton flux are deduced by considering a certain value for the IHS width $\Delta r$=30AU. The variability of the width does not affect the analysis and results of this study, in regards to the thermodynamic parameters of temperature, polytropic index, and entropy, as explained below:

(1) How much is the temperature and density affected by the choice of $\Delta r$?

The ENA flux is proportional to the radial average of the proton flux and the width of the inner heliosheath; in other words, for an observed ENA flux, the proton (radial average of) flux is inverse proportional to the width of the inner heliosheath. If the width characterizing the $i^{\text{th}}$ pixel were to change from $\Delta r_i^{\text{old}}$ to $\Delta r_i^{\text{new}}$, the corresponding change of the proton flux will be $\log J_{*i}^{\text{new}} = \log J_{*i}^{\text{old}} - \log(\Delta r_i^{\text{new}} / \Delta r_i^{\text{old}})$. Neighboring pixels are having similar variation of the width, so that we may rewrite the previous relationship as

$$\log J_{*i}^{\text{new}} \sim \log J_{*i}^{\text{old}} + \langle \log(\Delta r^{\text{new}} / \Delta r^{\text{old}}) \rangle, \quad (C1)$$

where the averaging applies to the pixels neighboring the $i^{\text{th}}$ pixel.

In the linear relationship between the values of the invariant $\Pi$ and kappa, the temperature is connected to the slope, while the density is connected to the intercept, as shown in Eq.(14), i.e.,

$$\log n_0 + (\kappa_i - \tfrac{1}{2}) \cdot \log(k_B T_0) = \log J_{*i} - \log f_{\kappa_i} = \log \Pi_i. \quad (C2)$$

with the subscript $i$ counting pixels with the same density and temperature (Section 3).

Then, writing Eq.(C2) for both old and new fluxes, we have,

$$\begin{aligned} \log n_0^{\text{old}} + (\kappa_i - \tfrac{1}{2}) \cdot \log(k_B T_0^{\text{old}}) &= \log J_{*i}^{\text{old}} - \log f_{\kappa_i}, \\ \log n_0^{\text{new}} + (\kappa_i - \tfrac{1}{2}) \cdot \log(k_B T_0^{\text{new}}) &= \log J_{*i}^{\text{new}} - \log f_{\kappa_i}, \end{aligned} \quad (C3)$$

and taking into account Eq.(C1), we have

$$\log n_0^{\text{new}} + (\kappa_i - \tfrac{1}{2}) \cdot \log(k_B T_0^{\text{new}}) \equiv \log n_0^{\text{old}} + \langle \log(\Delta r^{\text{new}} / \Delta r^{\text{old}}) \rangle + (\kappa_i - \tfrac{1}{2}) \cdot \log(k_B T_0^{\text{old}}), \quad (C4a)$$

or

$$\log n_0^{\text{new}} = \log n_0^{\text{old}} + \langle \log(\Delta r^{\text{new}} / \Delta r^{\text{old}}) \rangle \text{ and } \log(k_B T_0^{\text{new}}) = \log(k_B T_0^{\text{old}}). \quad (C4b)$$

Therefore, the logarithm of density is affected by the logarithm of the width ratio, $\log n_0^{\text{new}} - \log n_0^{\text{old}} = \langle \log(\Delta r^{\text{new}} / \Delta r^{\text{old}}) \rangle$, while the temperature is unaffected, $\log(k_B T_0^{\text{new}}) - \log(k_B T_0^{\text{old}}) = 0$.

(2) Moderate variations of thickness.

The (logarithm of) density is affected by $\log n_0^{\text{new}} - \log n_0^{\text{old}} = \langle \log(\Delta r^{\text{new}} / \Delta r^{\text{old}}) \rangle$, which is approximated by $\sim (\Delta r^{\text{new}} - \Delta r^{\text{old}}) / [\Delta r^{\text{old}} \ln(10)]$. An increase of the inner heliosheath thickness $\Delta r$~30 AU by ~5AU: would have impacted the density by $\Delta n/n$~15% and the density logarithm by $\Delta \log n$~0.065; even with a larger variation $\delta \Delta r$~20 AU (i.e., $\Delta r$~50AU), we obtain $\Delta \log n$~0.22; that is, quite smaller than the uncertainties of the derived (logarithm of) densities; indeed, the statistics of the histogram of the all-years density uncertainty, $\delta \log(n/\text{cm}^{-3})$, provides the following mean, mode, and median, respectively: 0.502, 0.246, and 0.316.

However, larger variations of thickness may affect the density significantly; for example, when comparing pixels near nose with pixels near tail, the width ratio may be larger than an order of magnitude, i.e., $\log n_0^{\text{new}} - \log n_0^{\text{old}} > 1$.

(3) Insignificant effect on polytropic index and entropy.

Polytropic index does not depend on the density but on the deviations of the logarithm of density. Any thickness variation would have affected similarly neighboring pixels, which correspond to subsequent values of density along the flow in the inner heliosheath. Therefore, thickness variation would have no impact on the polytropic index.

Moreover, the density is involved in the entropic formulation by its logarithm, and thus, any width or density variation has small impact on the value of entropy. For instance, an increase of the inner heliosheath thickness $\Delta r$~30 AU by ~5AU, would impact the entropy by $\Delta S/S$~4% or $\Delta \log S$~0.02.

**Appendix D. Method: Statistical Analysis**
**D.1. Description**

For each ($6^0 \times 6^0$) sky-map pixel, corresponding to a sky-map direction at $\Omega(\vartheta, \varphi)$, we perform the following statistical analyses (scheme in Figure D.1):

(i) Bi-parametrical linear fitting of the energy-flux spectrum to the dataset $\{\log \varepsilon_i(\Omega), \log j_i(\Omega) \pm \delta \log j_i(\Omega)\}$ for $i$=1, 2, …5, to derive the optimal values of the two fitting parameters $\{\kappa(\Omega) \pm \delta \kappa(\Omega), \log j_*(\Omega) \pm \delta \log j_*(\Omega)\}$, and from Eq.(13), they can be easily transformed to kappa-Π values, $\{\kappa(\Omega) \pm \delta \kappa(\Omega), \log \Pi(\Omega) \pm \delta \log \Pi(\Omega)\}$. (See D.2.1).

(ii) Smoothing of $\{\kappa(\Omega) \pm \delta \kappa(\Omega), \log \Pi(\Omega) \pm \delta \log \Pi(\Omega)\}$ and $\{\kappa(\Omega') \pm \delta \kappa(\Omega'), \log \Pi(\Omega') \pm \delta \log \Pi(\Omega')\}$, to derive the temperature and density sets $\{\log n(\Omega; \Omega') \pm \delta \log n(\Omega; \Omega')\}$ and $\{\log T(\Omega; \Omega') \pm \delta \log T(\Omega; \Omega')\}$, where $\Omega'$ counts any of the eight pixels that neighbor the pixel at $\Omega$. (See D.2.2).

(iii) A weighted averaging (mono-parametrical fitting) of the sets $\{\log T(\Omega; \Omega') \pm \delta \log T(\Omega; \Omega')\}$, $\{\log n(\Omega; \Omega') \pm \delta \log n(\Omega; \Omega')\}$ to estimate the optimal values $\log T(\Omega) \pm \delta \log T(\Omega)$, $\log n(\Omega) \pm \delta \log n(\Omega)$. (See D.2.3).

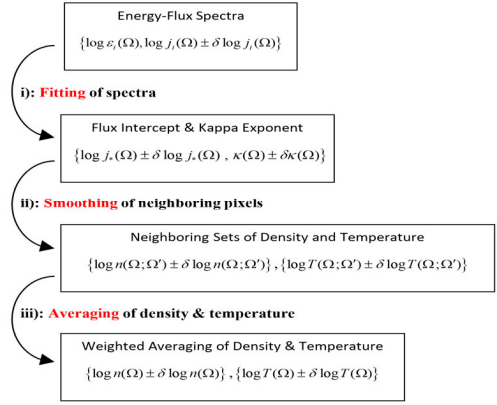

**Figure D.1.** The three main steps of the performed statistical analyses of the method.

### D.2. Statistical Analyses

*D.2.1. Fitting of energy – flux spectra*

We apply a linear fitting to the energy – flux spectra (on a log-log scale) characterizing the sky-map direction at Ω, in order to estimate the respective optimal values of the flux-logarithm intercept $\log j_*$ and the kappa $\kappa$ (according to Eq.(5)).

We fit the bi-parametrical linear statistical model of $y = p_1 + p_2 \cdot x$, with variables $(x = \log \varepsilon, y = \log j)$ and fitting parameters $(p_1 = \log j_*, p_2 = -\kappa)$, to the dataset $\{\log \varepsilon_i, \log j_i \pm \delta \log j_i\}$ for $i$=1,2,…5. We minimize the chi-square $\chi^2(p_1, p_2) = \sum_{i=1}^{N} W_i (\log j_i - p_1 - p_2 \log \varepsilon_i)^2$, for the number of $N$=5 IBEX-Hi ESA 2-6, where the weights $W_i = \tau_i / \sum_{i=1}^{N} \tau_i$ are given in terms of the exposure time of each energy channel $\tau_i$. The global minimum of the chi-square gives the optimal parameter values, $(p_1^*, p_2^*)$, by solving the equations $\partial \chi^2(p_1, p_2) / \partial p_1 = 0$ and $\partial \chi^2(p_1, p_2) / \partial p_2 = 0$.

The total error, characterizing each of the optimal values of the fitting parameters, is caused by two independent uncertainties: (i) the fitting error, which results from the nonzero square of the residuals, and (ii) the propagation error, which results from the propagation of the uncertainties characterizing the temperature values. The total error comes the summation of the respective variances (considering their independent source). (For more details, see: Bevington 1969; Livadiotis 2007; 2014b; 2018c).

The fitting errors of these values are given by $\delta p_{\alpha,\text{st}}^* = \sqrt{\chi_{\text{red}}^2 \cdot H_{\alpha\alpha}^{-1}}$, $\alpha$=1,2, where $H$ is the Hessian matrix of the chi-square at the global minimum, and $H_{\alpha\alpha}^{-1}$ is the $\alpha$-th diagonal element of its inverse matrix (Melissinos 1966; Livadiotis 2007); the estimated chi-square value is $\chi_{\text{est}}^2 = \chi^2(p_1^*, p_2^*)$, while $\chi_{\text{red}}^2 = \frac{1}{N-2} \chi_{\text{est}}^2$ is the reduced chi-square with degrees-of-freedom (DOF) $M$=$N$-2. (For details about $\chi_{\text{red}}^2$, $p$-value, and the goodness of fitting, see: Section 3). The propagation errors due to the flux uncertainties are $\delta p_{\alpha,\text{pr}}^* = \sqrt{\sum_{i=1}^{N} (\partial p_\alpha^* / \partial \log j_i)^2 (\delta \log j_i)^2}$, $\alpha$=1,2, with the derivatives calculated numerically.

*D.2.2. Smoothing of neighboring temperature and density values*

In general, the smoothing of the pairs of values $\{(x_1, y_1), (x_2, y_2), \cdots, (x_N, y_N)\}$ involves finding the best fit to a constant pair value $(x_{\text{opt}}, y_{\text{opt}})$. The minimization of the variance or the chi-square leads to the optimal value, the (weighted) mean. The smaller the least chi-square, the better the smoothing, while the perfect smoothing would occur when the pairs were already the same, i.e., $(x_1, y_1) = \ldots = (x_N, y_N)$.

Here we ask: how can we work in the case where each pair of values produces a specific solution or constraint? The constraint is expressed by an equation of $x$ and $y$, having the same form for each point. This can be generally written as: $F(x_1, y_1; a_1, b_1) = 0$, $F(x_2, y_2; a_2, b_2) = 0$, ..., $F(x_N, y_N; a_N, b_N) = 0$, or in a linear form, as $x_1 + a_1 y_1 = b_1$, $x_2 + a_2 y_2 = b_2$, and $x_N + a_N y_2 = b_N$, where $a$'s and $b$'s are known given numbers. Again, we wish to find the pair of optimal values $(x_{opt}, y_{opt})$ that lies closer to the whole set of pair values, that is, to minimize the respective chi-squares, $\sum_{i=1}^{N} \sigma_x^{-2} \cdot (x_i - x_{opt})^2$ and $\sum_{i=1}^{N} \sigma_y^{-2} \cdot (y_i - y_{opt})^2$ (where $\sigma_x$ and $\sigma_y$ are the errors of $x_i$ and $y_i$, respectively).

Alternatively, the smoothing can be performed by pair combinations. First, we solve all combinations of two equations, e.g., the $i^{th}$ and $j^{th}$ equations, $F(x, y; a_i, b_i) = 0$ and $F(x, y; a_j, b_j) = 0$, where we set $x_1 = x_2 = x$ and $y_1 = y_2 = y$, leading to the values of $x(a_i, b_i; a_j, b_j)$ and $y(a_i, b_i; a_j, b_j)$, e.g., for the linear examples: $x = (a_j b_i - a_i b_j)/(a_j - a_i)$, $y = (b_j - b_i)/(a_j - a_i)$. Then, the optimal values are derived by a mono-parametrical fitting minimizing the variance of the estimated $ij$ components, leading to their weighted average, $\sum_{i=1}^{N} \sigma_x^{-2} \cdot [x(a_i, b_i; a_j, b_j) - x_{opt}]^2$ and $\sum_{i=1}^{N} \sigma_y^{-2} \cdot [y(a_i, b_i; a_j, b_j) - y_{opt}]^2$. The goodness of the fitting (measured by the reduced chi-square and $p$-value) is decisive for the acceptance or not of the smoothing under a certain statistical confidence (see Section 3 and Figure 4).

Here we use the smoothing performed by pair combinations, in order to calculate the values of the temperature and density, which are derived for each one of the eight neighboring pixels (noted by $\Omega'$) surrounding the central pixel (noted by $\Omega$). This produces eight pairs of temperature and density values, all characterizing the central pixel at $\Omega$.

In particular, the estimation of density and temperature characterizing the direction $\Omega$, can be derived by combining the pixel at the direction $\Omega$ with any of the eight closest pixels with directions $\Omega'$, where $\Omega = (\vartheta, \varphi)$ and $\Omega' = (\vartheta + 6^0 \cdot i, \varphi + 6^0 \cdot j)$ with $i, j$: -1, 0, +1, and excluding $i=j=0$. Then, for each $6^0 \times 6^0$ sky-map pixel direction, we derive the sets $\{\log n(\Omega; \Omega') \pm \delta \log n(\Omega; \Omega')\}$ and $\{\log k_B T(\Omega; \Omega') \pm \delta \log k_B T(\Omega; \Omega')\}$. The smoothing analysis combines Eq.(14) for neighboring pixels, i.e., Eq.(15). Hence, we end up with Eq(16), or substituting $\Pi = J_* / f_\kappa$ from (13) or (14),

$$\log n = (\kappa - \kappa')^{-1} \cdot [(\kappa' - \tfrac{1}{2}) \cdot (\log J_* - \log f_\kappa) - (\kappa - \tfrac{1}{2}) \cdot (\log J_*' - \log f_{\kappa'})], \tag{D1a}$$

$$\log(k_B T) = (\kappa - \kappa')^{-1} \times (\log J_* - \log J_*' - \log f_\kappa + \log f_{\kappa'}), \tag{D1b}$$

where we have set $\kappa \equiv \kappa(\Omega)$, $\kappa' \equiv \kappa(\Omega')$, $\log J_* \equiv \log J_*(\Omega)$, $\log J_*' \equiv \log J_*(\Omega')$.

The corresponding errors are estimated using the propagation of the errors of the statistically independent parameters kappa $\kappa$ and flux-logarithm intercept $\log J_*$, which are involved in Eqs.(D1a,b):

$$\delta \log n = I_n \cdot (\kappa - \kappa')^{-2}, \quad \delta \log(k_B T) = I_T \cdot (\kappa - \kappa')^{-2}, \tag{D2}$$

where we have set:

$$I_T^2 \equiv (\kappa - \kappa')^2 [(\delta \log J_*)^2 + (\delta \log J_*')^2]$$
$$+ [\log f_\kappa - g(\kappa)(\kappa - \kappa') / \ln 10]^2 \delta \kappa^2 + [\log f_{\kappa'} + g(\kappa')(\kappa - \kappa') / \ln 10]^2 \delta \kappa'^2, \tag{D3}$$

$$I_n^2 \equiv (\kappa - \kappa')^2 \left[ (\kappa' - \tfrac{1}{2})^2 (\delta \log J_*)^2 + (\kappa - \tfrac{1}{2})^2 (\delta \log J_*')^2 \right] +$$
$$\left[ \log J_*' + \log f_\kappa - g_\kappa (\kappa - \kappa') / \ln 10 \right]^2 (\kappa' - \tfrac{1}{2})^2 \delta \kappa^2 + \left[ \log J_* + \log f_{\kappa'} + g_\kappa (\kappa - \kappa') / \ln 10 \right]^2 (\kappa - \tfrac{1}{2})^2 \delta \kappa'^2, \tag{D4}$$

$$g_\kappa \equiv d \ln f_\kappa / d\kappa = \ln(\kappa - \tfrac{3}{2}) + 1 + (\kappa - \tfrac{3}{2})^{-1} + \Psi(\kappa + 1) - \Psi(\kappa - \tfrac{1}{2}), \tag{D5}$$

(where $\Psi$ denotes the digamma function).

*D.2.3. Averaging of the temperature and density values of neighboring pixels*

We apply a fitting technique to the derived sets of temperature $\{\log T(\Omega;\Omega') \pm \delta \log T(\Omega;\Omega')\}$ and density $\{\log n(\Omega;\Omega') \pm \delta \log n(\Omega;\Omega')\}$, in order to estimate the respective weighted average values characterizing the sky-map direction at $\Omega$, $\log T(\Omega) \pm \delta \log T(\Omega)$ and $\log n(\Omega) \pm \delta \log n(\Omega)$.

In general, the fitting of the mono-parametrical statistical model $\{V(x_i;c)\}_{i=1}^N$ to a given dataset of $N$ $y$-values, $\{y_i\}_{i=1}^N$, involves finding the optimal parameter value $c = c^*$ that minimizes the total square deviations between model and data. Here, we just focus on the details of fitting of temperature data, while the density data are addressed likewise.

In particular, in order to find the optimal values of the (logarithm of) temperature, $\log T(\Omega) \pm \delta \log T(\Omega)$, the fitting involves minimizing the chi-square $\chi^2(c) = \sum_{i=1}^{8} \delta \log T_i^{-2} (\log T_i - c)^2$, where the set of fitted data is $\log T_i \pm \delta \log T_i$, for $i=1, ..., N=8$, while the statistical model used is simply given by the parameter, $V(t;c) = c$; namely, we fit a constant function to the set of eight data points. The resulted optimal value is given by the weighted mean $c^* = \sum_{i=1}^{8} w_i \log T_i$, where the weight is $w_i = \delta \log T_i^{-2} / \sum_{i=1}^{8} \delta \log T_i^{-2}$. The minimized chi-square, $\chi^2(c^*) = \sum_{i=1}^{8} \delta \log T_i^{-2} (\log T_i - c^*)^2$, can be used to characterize the goodness of the fitting (see further below).

The total error $\delta c^*$, characterizing the optimal value $c^*$ of the fitting parameter $c$, is again produced by two independent uncertainties: (i) the fitting error, which results from the nonzero square of the residuals and equals the standard error of the mean, $\delta c_{st}^* = [B^{-1} \cdot \sum_{i=1}^{8} w_i (\log T_i - c^*)^2]^{\frac{1}{2}}$, where the factor $B = (\sum_{i=1}^{8} w_i^2)^{-1} - 1$ is called effective DOF and reduces the bias caused by the usage of the statistical weights, and (ii) the propagation error, which results from the propagation of the uncertainties characterizing the temperature values, $\{\delta \log T_i\}_{i=1}^{8}$ and equals $\delta c_{pr}^* = (\sum_{i=1}^{8} \log T_i^{-2})^{-\frac{1}{2}}$. The total variance sums these variances (considering their independent source), i.e., $\delta c^* = (\delta c_{st}^{*2} + \delta c_{pr}^{*2})^{\frac{1}{2}}$. (For more details on the error analysis, see: Bevington 1969; Livadiotis 2007; 2014b; 2018c).

Next, we address the question, whether the smoothing and averaging is statistically confident, namely, the sets $\{\log T(\Omega;\Omega') \pm \delta \log T(\Omega;\Omega')\}$ and $\{\log n(\Omega;\Omega') \pm \delta \log n(\Omega;\Omega')\}$ are well-fitted by a constant, and thus, well-represented by their mean value. While we can always assign an average value to a series of numbers, whether this is a confident representative is a matter of statistics.

For the smoothing, we examine the null hypothesis ($H_0$): no (statistically significant) variations of the temperature and density values of the proton population in the IHS, between two neighboring sky-map $6^0 \times 6^0$-pixel directions. Practically, this entails that the sets $\{\log T(\Omega;\Omega') \pm \delta \log T(\Omega;\Omega')\}$ and $\{\log n(\Omega;\Omega') \pm \delta \log n(\Omega;\Omega')\}$ can be well represented by their (direction-$\Omega'$ independent) weighted mean and error, $\log T(\Omega) \pm \delta \log T(\Omega)$ and $\log n(\Omega) \pm \delta \log n(\Omega)$.

The null hypothesis is tested by examining the estimated chi-square minimum value of the involved mono-parametrical fitting; then, the goodness of this fitting is determined according to the following two common ways: (a) the *p*-value of the extremes, that is, the probability of having an extremer chi-square value, must be larger than a certain confidence level; and, (b) the reduced chi-square value has to be smaller as possible, but preferably closer to ~1. More analytically, we have:

(a) *p-value of the extremes*. We estimate the goodness of the fitting by comparing the derived chi-square minimum with the chi-square distribution (i.e., the distribution of all the chi-square minima),

$$P(\chi^2; M) d\chi^2 = 2^{-\frac{1}{2}M} \Gamma(\tfrac{1}{2}M)^{-1} \cdot (\chi^2)^{\frac{1}{2}M-1} \cdot e^{-\frac{1}{2}\chi^2} d\chi^2. \tag{D6}$$

The likelihood of having a chi-square value smaller than the estimated value $\chi^2_{est}$, is given by the cumulative distribution $P(0 \leq \chi^2 \leq \chi^2_{est}) = \int_0^{\chi^2_{est}} P(\chi^2; M) d\chi^2$, while the likelihood of having a chi-square value larger than the estimated value $\chi^2_{est}$, is given by the complementary cumulative distribution $P(\chi^2_{est} \leq \chi^2 < \infty) = \int_{\chi^2_{est}}^{\infty} P(\chi^2; M) d\chi^2$. Therefore, the probability of taking a result extremer than the observed value determines the *p*-value, i.e., the smallest between $P(0 \leq \chi^2 \leq \chi^2_{est})$ and $P(\chi^2_{est} \leq \chi^2 < \infty)$. A null hypothesis associated with a *p*-value smaller than a confidence level is typically rejected. (Commonly used confidence levels are: 0.01 and 0.05.)

(b) *Reduced Chi-Square value*. The reduced chi-square refers to the minimum chi-square value, normalized to DOF $M = 7$, i.e., $\chi^2_{red} = \tfrac{1}{7} \chi^2(c^*)$; to include the bias-correction, this becomes $\chi^2_{red} = \tfrac{B}{8} \sum_{i=1}^{8} \log T_i^{-2} (\log T_i - c^*)^2$, or written as $\chi^2_{red} = \delta c_{st}^{*2} / \delta c_{pr}^{*2}$. (The bias-reduction factor becomes $B=7$ when all the weights $w_i$ or errors $\log T_i$ are equal.) The fitting is "good" when $\chi^2_{red} \sim 1$. The reasoning behind this condition is that the standard error of the mean, i.e., the fitting error $\delta c_{st}^*$, is desired to be smaller and closer to zero as possible, but it is meaningless to be smaller than the data uncertainties, on average, that is, the propagation error $\delta c_{pr}^*$. The best scenario is when the data standard error is as small as the propagation error, $\delta c_{st}^* \sim \delta c_{pr}^*$. Given $\delta c_{st}^{*2} = \chi^2_{red} \cdot \delta c_{pr}^{*2}$, the condition $\delta c_{st}^* \sim \delta c_{pr}^*$ becomes $\chi^2_{red} \sim 1$.

In general, cases with $\delta c_{st}^* \gg \delta c_{pr}^*$ or $\chi^2_{red} \gg 1$ are not accepted. Here, we collect all cases with statistical variance equal ~3 times the propagation variance, or less, i.e., $\chi^2_{red} \leq 3$. Small values of $\chi^2_{red}$ (<< 1) correspond to small *p*-values (<< 0.05), but they also point to a small sum of residuals or standard deviation between the eight fitted data points and their estimated mean. These cases can be conditionally accepted, namely, the estimated mean temperature must have small uncertainty (error of the mean). Practically, we accept temperatures with error smaller than half order of magnitude, i.e., $\delta \log T \leq 0.5$. Figure 3 shows an example, where we consider the condition of small reduced chi-square (together with small relative error) $\chi^2_{red} < 3$ and $\delta \log T \leq 0.5$, instead of *p*-value > 0.01.

Finally, we provide the overall distinction of the pixels according to the goodness of the averaging, i.e., the fitting that provides the smoothed temperature and densities of neighboring pixels.

(a) *Pixels having invariant smoothed temperature and density but variable kappa, when compared with neighboring pixels*. The condition that approves for averaging the eight values of temperature and density, is to have sufficiently small reduced chi-square ($\chi^2_{red} \leq 3$) and relative temperature error ($\delta \log T \leq 0.5$), leading to an acceptable statistical confidence of the null statistical hypothesis that neighboring pixels have the same smoothed temperature and density.

(b) *Pixels having variable density, temperature, and kappa, with neighboring pixels*. The averaging of the eight values of temperature and density, is not statistically confident, if $\chi^2_{red} > 3$. Then, the density and temperature values of neighboring pixels vary significantly and cannot be smoothed.

(c) *Pixels having invariant kappa with their neighboring pixels*. Not all the eight neighboring pixels are equally contributing to the fitting for deriving the average temperature (and density). Some of the temperature (and density) values may have large uncertainties, caused by the small difference between the two kappa values $|\kappa(\Omega) - \kappa(\Omega')| \leq \delta\kappa = \sqrt{\delta\kappa(\Omega)^2 + \delta\kappa(\Omega')^2}$. Finally, if $|\kappa(\Omega) - \kappa(\Omega')| \geq \delta\kappa$ for at least six out of the eight points, then the kappa value of the central pixel is considered to remain invariant with the kappa of its neighbors, and thus, the density or temperature cannot be determined for this pixel.

The weighted mean of the kappa values of the eight neighboring pixels, for which $|\kappa(\Omega) - \kappa(\Omega')| \leq \delta\kappa$, provides a "smoothed" value of kappa, $\kappa_{sm}(\Omega)$, and is expected to be close enough to the originally derived value of kappa of the central pixel, $\kappa(\Omega)$; namely, $|\kappa(\Omega) - \kappa_{sm}(\Omega)| \leq \sqrt{\delta\kappa(\Omega)^2 + \delta\kappa_{sm}(\Omega)^2}$, otherwise, the smoothing of kappa is rejected.

(d) *Pixels with kappa less than 1.5*. Hard spectral indices corresponding to $\kappa < 1.5$ are not allowed in the formulation of kappa distributions. These spectral indices may be caused by the following: (i) when protons dominate in low-energy regions, the high-energy condition, $\varepsilon \gg (\kappa - \frac{3}{2})k_B T$ and $\varepsilon \gg \varepsilon_b$, is not sufficiently fulfilled (e.g., Zirnstein et al. 2021b studied the impact of the 3D isotropic turbulence on energy-flux spectra); (ii) newly born pickup protons, which have not been incorporated in the proton plasma, lead to a curved spectral tail that deviates from the linear behavior (on a log-log scale) (e.g., Livadiotis et al. 2012 studied the impact of newly born pickup protons and the proton plasma flow on spectra); and, (iii) the existence of more than one proton population described by a superposition of kappa distributions (e.g., Dayeh et al. 2012, studied the impact of superposition of the bi-modal - fast and slow - solar wind, exhibiting stronger effect in spectra near polar regions; see also: Desai et al. 2019).

Figure 6 in Section 3 plots the annual percentages of each of the above categories for each of the observational years between 2009 and 2016.

**Appendix E. Entropic formulation associated with kappa distributions**

The entropy associated with kappa distributions is explicitly expressed in terms of the density, temperature, and kappa. The formulation was derived in (Livadiotis 2018b; Livadiotis & McComas 2021), generalizing the particle entropy of Sackur (1911) and Tetrode (1912), that is,

$$S = \kappa \cdot [1 - (T/T_{S0})^{-\frac{3}{2}\cdot\frac{1}{\kappa}}], \qquad (E1)$$

where the thermal constant $T_{S0}$ constitutes the minimum temperature for the entropy to be positive (Livadiotis & McComas 2013c; Livadiotis 2014a; 2017, Chapters 2 and 5),

$$k_B T_{S0} = C \cdot \hbar_c^2 (m_e m_i)^{-\frac{1}{2}} \cdot \lambda_c^{-2} \cdot g_{\kappa,N}, \qquad (E2)$$

where $C = (9\pi/2)^{1/3}/e \approx 0.89$; $\lambda_c$ constitutes the smallest correlation length, interpreted by the interparticle distance $b \sim n^{-1/d}$ for collisional particle systems (absence of correlations), or by the Debye length for collisionless particle systems (where long-range interactions induce correlations among particles). The factor $g_{\kappa,N}$ depends on kappa and the number of correlated particles $N$, with $g_{\kappa,N} \approx 1$ for large $N$; the phase-space parcel $\hbar_c$ is typically given by the Planck's constant, but it was shown to represent a different and larger constant in space plasmas, where Debye shielding limits the distance of correlations, $\hbar_* = (1.19 \pm 0.05) \times 10^{-22} \text{ J} \cdot \text{s}$ (Livadiotis & McComas 2013c; 2014; Livadiotis & Desai 2016; Livadiotis 2016; 2019c; Livadiotis et al. 2020); i.e., $\hbar_c = \hbar$ in the absence of correlations, while $\hbar_c = \hbar_*$ when correlations exist among particles beyond the nearest neighbors, as for the majority of space plasmas.